\documentclass[usenatbib]{mn2e}
\newcommand{\subC}{_{_{\rm CENTRE}}}

\newcommand{\subD}{_{_{\rm DUST}}}
\newcommand{\subF}{_{_{\rm FLOW}}}

\newcommand{\subH}{_{_{{\rm H}_2{\rm .FLOW}}}}
\newcommand{\subL}{_{_{\rm LAYER}}}
\newcommand{\subO}{_{_{\rm O}}}
\newcommand{\subS}{_{_{\rm SHOCKED}}}

\usepackage{graphicx,tabularx,amsmath,amssymb,upgreek,wasysym,mathtools,multirow}
\numberwithin{equation}{section}
\title[A ram-pressure threshold for star formation]{A ram-pressure threshold for star formation}
\author[A. P. Whitworth]{A. P. Whitworth\thanks{E-mail: ant at astro.cf.ac.uk}\\
School of Physics and Astronomy, Cardiff University, Cardiff CF24 3AA, Wales, UK}

\begin{document}
\pagerange{\pageref{firstpage}--\pageref{lastpage}} \pubyear{2013}
\maketitle
\label{firstpage}

\begin{abstract} 
In turbulent fragmentation, star formation occurs in condensations created by converging flows. The condensations must be sufficiently massive, dense and cool to be gravitationally unstable, so that they start to contract; {\it and} they must then radiate away thermal energy fast enough for self-gravity to remain dominant, so that they continue to contract. For the metallicities and temperatures in local star forming clouds, this second requirement is only met robustly when the gas couples thermally to the dust, because this delivers the capacity to radiate across the full bandwidth of the continuum, rather than just in a few discrete spectral lines. This translates into a threshold for vigorous star formation, which can be written as a minimum ram-pressure $P_{_{\rm CRIT}}\!\sim\! 4\!\times\! 10^{-11}{\rm dyne}$. $P_{_{\rm CRIT}}$ is independent of temperature, and corresponds to flows with molecular hydrogen number-density $n_{_{{\rm H}_2{\rm .FLOW}}}$ and velocity $v\subF$ satisfying $n_{_{{\rm H}_2{\rm .FLOW}}} v\subF^2\ga 800\,{\rm cm}^{-3}({\rm km}/{\rm s})^2$. This in turn corresponds to a minimum molecular hydrogen column-density for vigorous star formation, $N_{_{{\rm H}_2{\rm .CRIT}}}\!\sim\! 4\!\times\! 10^{21}{\rm cm}^{-2}$ ($\Sigma_{_{\rm CRIT}}\!\sim\! 100\,{\rm M}_{_\odot}{\rm pc}^{-2}$), and a minimum visual extinction $A_{_{\rm V,CRIT}}\!\sim\! 9\,{\rm mag}$. The characteristic diameter and line-density for a star-forming filament when this threshold is just exceeded -- a {\it sweet spot} for local star formation regions -- are $2R_{_{\rm FIL}}\!\sim\! 0.1{\rm pc}$ and $\mu_{_{\rm FIL}}\!\sim\! 13{\rm M}_{_\odot}{\rm pc}^{-2}$. The characteristic diameter and mass for a prestellar core condensing out of such a filament are $2R_{_{\rm CORE}}\!\sim\! 0.1{\rm pc}$, and $M_{_{\rm CORE}}\!\sim\! 1\,{\rm M}_{_\odot}$. We also show that fragmentation of a shock-compressed layer is likely to commence while the convergent flows creating the layer are still ongoing, and we stress that, under this circumstance, the phenomenology and characteristic scales for fragmentation of the layer are fundamentally different from those derived traditionally for pre-existing layers.
\end{abstract}


\section{Introduction}\label{SEC:INTRO}

It is now widely believed that stars form in molecular clouds, as a consequence of turbulent fragmentation \citep[e.g.][]{Elmegree2000,PadoNord2002,HennChab2009}. A range of mechanisms, from the inflows assembling a molecular cloud, to outflows from the stars that have already formed inside it, inject turbulent energy into the cloud \citep[e.g.][]{Balletal2007}. At certain places, the convergent flows generated by the turbulence are sufficiently strong and coherent to produce dense condensations. Whether a condensation is prestellar (i.e. whether it condenses out to form at least one star, and quite often a small group of stars) depends {\it both} on whether it is sufficiently massive and cool for self-gravity to dominate over internal pressure (so that the condensation starts to contract), {\it and} on whether it can stay sufficiently cool for self-gravity to remain dominant over internal pressure (so that contraction of the condensation is not reversed by internal pressure). Specifically, the condensation must be able to radiate away immediately a large fraction of the gravitational potential energy being released by contraction \citep[e.g.][]{Rees1976,LowCLynd1976}. Otherwise, the gas heats up due to compression, and the contraction is either reversed in an `adiabatic bounce', or slowed sufficiently that the condensation is likely to merge with neighbouring condensations or be sheared apart.

In local star-forming molecular clouds, the gas-kinetic temperature is typically $T\!\sim\!10\,{\rm K}$ (but in some extreme locations it may rise to $T\!\sim\!40\,{\rm K}$), and there are two dominant channels for radiative cooling: line emission from gas particles, and continuum emission from dust grains. Line emission from gas particles entails the collisional excitation of ions, atoms and molecules (at the expense of the thermal kinetic energy of the gas) followed by spontaneous radiative de-excitation and the emission of line photons that then escape from the cloud. In this regard, key gas-phase species are C$^+$ ions, C$^0$ and O$^0$ atoms, and CO molecules \citep[e.g.][]{DalgMcCr1972,HollMcKe1979,Wolfetal1995}. Continuum emission from the dust entails gas particles colliding with, becoming adsorbed on, and hence delivering their thermal kinetic energy to, dust grains. The dust grains then radiate the energy away in the continuum \citep{HollMcKe1979}. If the gas particles are subsequently evaporated from the grain surface, they normally leave with kinetic energy characteristic of the vibrational temperature of the dust grain, $T\subD$. Consequently, if $T\subD\!<\!T$, there has been a net loss of thermal kinetic energy from the gas.

In this paper we explore the possibility that there is a threshold, or sweet spot, for star formation when the density becomes sufficiently high, and the rate of transfer of thermal energy from the gas to the dust sufficiently fast, that the dominant cooling channel switches from line emission by gas particles to continuum emission by dust. The resulting abrupt increase in the bandwidth available for cooling delivers self-gravitating condensations into a regime where they can radiate much more efficiently. Instead of radiating in a few narrow emission lines --- many of which are by this stage becoming optically thick, and/or the molecules involved are freezing out onto the dust ---  self-gravitating condensations can now radiate over the entire bandwidth of the blackbody spectrum. If this hypothetical threshold is indeed important, the principal locations where star formation occurs are the places where convergent turbulent flows have assembled a sufficient mass of gas that is {\it both} dense enough to be gravitationally unstable, {\it and} dense enough to stay cool by transferring its thermal energy to the dust. These requirements translate into a critical ram pressure in the turbulent flows creating the condensation, a critical column-density (and hence a critical dust extinction), a critical length-scale, a critical line density, and a critical mass.

In Section \ref{SEC:BASICS} we derive the basic results, using very simple arguments. The rest of the paper deals with a more analytic discussion, describing the basic features of an idealised model of turbulent fragmentation and the underlying thermodynamics; it should be skipped by those not concerned with such details. In Section \ref{SEC:DYNAMICS} we review the dynamics of the formation of dense shock-compressed layers, and we derive the length- and mass-scales characterising the gravitational fragmentation of such layers into filaments, and then into cores, {\it assuming} that the gas flowing into the layer is able to cool effectively by emitting radiation; the basic expression for the gravitational acceleration driving the growth of corrugation instabilities in a layer, and the time evolution of the cylindrical filaments that form from such corrugation instabilities, are derived in Appendices \ref{APP:gY} and \ref{APP:CYL}. In Section \ref{SEC:THERMO} we present approximate analytic expressions for the thermal properties of star forming gas and for the cooling rates due to CO line emission by the gas, and continuum emission by the dust; these cooling rates are derived in Appendices \ref{APP:CO} and \ref{APP:DUST}. We use these expressions to evaluate the circumstances under which (i) the gas flowing into a shock compressed layer is able to radiate effectively, so that the bulk of the layer is cool and can fragment (first into filaments and then into cores), showing that molecular line radiation is likely to be the dominant cooling channel for the gas immediately behind the accretion shock bounding such a layer; (ii) the cores formed by fragmentation of a shock-compressed layer are able to radiate fast enough to condense out as protostars, showing that dust cooling is likely to be the dominant cooling channel in a condensing prestellar core. We note that, between the gas flowing into the layer and cooling down by molecular line radiation (circumstance (i) above), and the gas participating in the collapse and fragmentation of a core and keeping cool by dust continuum radiation (circumstance (ii) above), there are lateral flows within the layer, collecting the gas into filaments, and then along the filaments into cores, but these flows occur at approximately constant density, and therefore do not depend critically on the capacity of the gas to cool radiatively. We discuss the consequences of this model of star formation, and summarise our main conclusions, in Section \ref{SEC:CONC}.

In the model developed here, the dynamics of turbulent fragmentation proceeds in three phases. First, colliding flows produce cold shock compressed layers; the cooling behind the accretion shocks bounding these layers is dominated by molecular line emission. Second, the shock-compressed layers fragment into filaments. This occurs at approximately constant density, so cooling is not an issue during this phase. It involves a fragmentation mode -- which we call the {\it early} fragmentation mode -- in which the material going into a filament is gathered from a region that is much more extended in the plane of the layer than the layer thickness -- and hence the filaments are quite well separated.\footnote{This early fragmentation mode is fundamentally different from the modes analysed by \citet{LarsonRB1985}, which we refer to as {\it late} fragmentation modes, for reasons explained in Section \ref{SEC:FundDiff}. It relates more closely to the analysis of \citet{ElmeElme1978}, in the sense that they do consider the effects of external pressure, and also the issue of whether the layer has sufficient time to relax to hydrostatic equilibrium. However, they do not explicitly relate the external ram pressure to the rate of growth of the layer, or the fact that its subsequent fragmentation should proceed at approximately constant density. These are fundamental elements of our model.} Third, prestellar cores condense out of the filaments, particularly at their intersections if there is no preferred orientation for the filaments. Core collapse and fragmentation, to form protostars, requires efficient cooling, and therefore proceeds much more vigorously if the density is high enough for the gas to couple thermally to the dust.

We stress that, in the analysis that follows, the purely numerical factors are very approximate and should not be given great weight. They are sometimes given to three significant figures to reduce the likelihood of cumulative shifts in quantities involving several such factors, but final numbers are then normally reduced to one significant figure. The inferences that we wish to stress are: (a) the systematic trends that are predicted; (b) the role of the ram pressure, $\rho\subF v\subF^2$, of the gas flowing into the layer with density $\rho\subF$ and velocity $v\subF$, since this is the critical factor that controls the density in the layer (and hence in the filaments and embryonic cores that condense out of it) and thereby determines whether the gas in the layer (or filament, or embryonic core) can  couple thermally to the dust; (c) the factors involving powers of the Mach Number, ${\cal M}=v\subF/a\subL$ (the ratio of the velocity of the matter flowing into a layer, $v\subF$, to the effective isothermal sound speed in the layer, $a\subL$), since it is these factors that distinguish, fundamentally, fragmentation of an {\it accumulating, shock-compressed}  layer from fragmentation of an {\it already accumulated} layer  \citep[as analysed, for example, by][see Sections \ref{SEC:FundDiff} and \ref{SEC:ReduDime} below]{LarsonRB1985}.

Throughout the paper we will formulate most equations in terms of mass-density $\rho$ and isothermal sound speed $a$, rather than molecular hydrogen number-density $n_{_{{\rm H}_2}}$ and temperature $T$, except where representative values are deemed easier to comprehend when expressed in terms of $n_{_{{\rm H}_2}}$ and/or $T$. We can switch between the two using
\begin{eqnarray}
\rho&=&n_{_{{\rm H}_2}}\bar{m}_{_{{\rm H}_2}}\,,\\
a&=&\left(\frac{k_{_{\rm B}}T}{\bar m}\right)^{1/2}\,.
\end{eqnarray}
where ${\bar m}_{_{{\rm H}_2}}$ is the mass associated with one hydrogen molecule and ${\bar m}$ is the mean gas-particle mass. For the purpose of making estimates, we assume that the gas has solar composition, viz. $X=0.70$ (fraction by mass of hydrogen), $Y=0.28$ (helium), and $Z=0.02$ (metallicity, including dust); that the fraction of dust by mass is $Z\subD=0.01$; that the hydrogen is entirely molecular; and that in dense gas the default temperature is $T\sim 10\,{\rm K}$. Hence, ${\bar m}_{_{{\rm H}_{2}}}\!=\!2m_{\rm p}/X\!=\!4.8\times 10^{-24}\,{\rm g}\,$ (where $m_{\rm p}$ is the proton mass), $\bar{m}\!=\!m_{\rm p}(X/2+Y/4+Z/30)^{-1}\!=\!4.0\times 10^{-24}\,{\rm g}$, and in dense gas the default isothermal sound speed is $a\sim 0.2\,{\rm km}\,{\rm s}^{-1}$. Since we often deal with ram pressures rather than hydrostatic pressures, we sometimes divide the pressure by $\bar{m}_{_{{\rm H}_2}}$, so that the resulting quantity is the product of a molecular hydrogen number-density and a squared velocity (see, for example, Eq. \ref{EQN:PCrit2}).

\section{The critical ram-pressure and the early fragmentation mode}\label{SEC:BASICS}

In this section we preempt the main results of the paper with simple evaluations of the critical quantities.

\subsection{Thermal coupling between gas and dust}

The rate per unit volume at which the gas transfers thermal energy to the dust (which, in the regime with which we are concerned, is essentially the same as the rate at which the gas is cooled by the dust) is of order\footnote{To avoid very long equations, coefficients like the one preceding $\rho^2a^3$ in Eq. (\ref{EQN:G2D}) will normally be written using the shorthand $[{\rm Eq\,E1}]_{_{\rm G2D}}$, where the content of the square bracket identifies the equation where the coefficient is fully defined, here Eq. (E1) in Appendix E, and the subscript identifies the role that the coefficient plays, here quantifying the rate at which thermal energy is transferred from the gas to the dust, i.e. {\tiny G2D}.}
\begin{eqnarray}\label{EQN:G2D}
\Lambda_{_{\rm G2D}}&\sim&[1.12\times 10^2\,{\rm cm}^2\,{\rm g}^{-1}]_{_{\rm G2D}}\,\rho^2\,a^3\,.
\end{eqnarray}
A detailed derivation of this result is given in Appendix \ref{APP:DUST}.

The rate per unit volume at which the gas in a collapsing prestellar core is heated by compression is of order
\begin{eqnarray}
\Gamma_{_{\rm COMP}}\!&\!\sim\!&\![7.14\times 10^{-4}\,{\rm cm}^{3/2}\,{\rm g}^{-1/2}\,{\rm s}^{-1}]_{_{\rm COMP}}\,\rho^{3/2}\,a^2.
\end{eqnarray}
A detailed derivation of this result is given in Section \ref{SEC:COREHEAT}.

The two rates are equal when the density is of order
\begin{eqnarray}\label{EQN:rhoCRIT}
n_{_{{\rm H}_2.{\rm CRIT}}}&\sim&2\times 10^4\,{\rm cm}^{-3}\,\left(\frac{T}{10\,{\rm K}}\right)^{-1}\,.
\end{eqnarray}
Our hypothesis is that this is a critical density above which the gas in a core couples so well to the dust, thermally, that the core can easily stay cool and condense to much higher densities. It is a minimum density, but, since the capacity of turbulence to deliver cores of ever higher density (and hence ever lower mass) declines quite rapidly, and requires the gas to pass through this density, we expect there to be a preference for prestellar cores having parameters corresponding approximately to this critical density.

\subsection{Critical parameters}

The ram pressure required to reach this critical density is given by
\begin{eqnarray}\nonumber
P_{_{\rm CRIT}}\;\,\equiv\;\,\rho\subF\,v\subF^2&\sim&\rho_{_{\rm CRIT}}\,a^2\\\label{EQN:PCrit1}
                                                                           &\sim&4\times 10^{-11}\,{\rm dyne},
\end{eqnarray}
or equivalently,
\begin{eqnarray}\label{EQN:PCrit2}
\frac{P_{_{\rm CRIT}}}{\bar{m}_{_{{\rm H}_2}}}\;\,\equiv\;\,n_{_{{\rm H}_2{\rm .FLOW}}}\, v\subF^2&\sim&800\,{\rm cm}^{-3}\,({\rm km}/{\rm s})^2\,\!.
\end{eqnarray}
We note that $P_{_{\rm CRIT}}$ is independent of temperature.

With this external pressure, gravitational fragmentation occurs when the surface-density is 
\begin{eqnarray}\label{EQN:SigmaCrit}
\Sigma_{_{\rm CRIT}}&\sim&\left(\!\frac{2P_{_{\rm CRIT}}}{\pi G}\!\right)^{1/2}\;\,\sim\;\,100\,{\rm M}_{_\odot}\,{\rm pc}^{-2}\,.
\end{eqnarray}
The corresponding column-density of molecular hydrogen is
\begin{eqnarray}\label{EQN:NH2Crit}
N_{_{{\rm H}_2.{\rm \rm CRIT}}}&=&\frac{\Sigma_{_{\rm CRIT}}}{\bar{m}_{_{{\rm H}_2}}}\;\,\sim\;\,4\times 10^{21}\;{\rm H}_2\;{\rm cm}^{-2}\,,
\end{eqnarray}
and the corresponding visual extinction is
\begin{eqnarray}\label{EQN:AVCrit}
A_{_{\rm V.CRIT}}&=&\frac{5\Sigma_{_{\rm CRIT}}\kappa\subD Q_{_{\rm V}}}{\ln(100)}\;\,\sim\;\,9\,{\rm mag.}
\end{eqnarray}
Here $\kappa\subD$ is the net geometric cross-section of all the dust grains in unit mass of gas and dust (see Eq.  \ref{EQN:kappaD2}); $\;Q_{_{\rm V}}\sim 1.5$ is the extinction efficiency of a representative dust grain at visual wavelengths; and the factor $\;5/\ln(100)=1.086\;$ converts optical-depth into magnitudes. The diameters and masses of representative cores are obtained by substituting $\bar{\rho}=\rho_{_{\rm CRIT}}$ in the equations for a critical Bonnor-Ebert sphere (Eqs. \ref{EQN:RBEM} and \ref{EQN:MBEM}),
\begin{eqnarray}\label{EQN:2RCrit}
2R_{_{\rm CORE}}&\sim&0.1\,{\rm pc}\,\left(\frac{T}{10\,{\rm K}}\right)\,; \\\label{EQN:MCrit}
M_{_{\rm CORE}}&\sim&1\,{\rm M}_{_\odot}\,\left(\frac{T}{10\,{\rm K}}\right)^2\,.
\end{eqnarray}
$2R_{_{\rm CORE}}$ is also, approximately, the predicted diameter of a filament (see Section \ref{SEC:TOY1}).

These values of $\Sigma_{_{\rm CRIT}}$, $N_{_{{\rm H}_2,{\rm CRIT}}}$, $A_{_{\rm V,CRIT}}$, $2R_{_{\rm CORE}}$ and $M_{_{\rm CORE}}$ accord quite well with observations of nearby low-mass star formation regions. In particular, there is observational evidence for a similar characteristic size and mass for filaments and cores in low-mass star formation regions \citep[e.g.][]{Konyetal2010,Arzoetal2011}. And it has been suggested that there might be a similar extinction or column-density threshold for efficient star formation \citep[e.g.][]{Johnetal2004,Heidetal2010,Ladaetal2010} -- although \citet{BurkHart2013} offer an alternative explanation for the observations in terms of underlying selection effects.

\subsection{The fundamental difference between fragmentation of an accumulating layer and fragmentation of an already accumulated layer}\label{SEC:FundDiff}

In the situation that we consider here there is a trade-off between two timescales: the timescale on which the shock-compressed layer accumulates, and the timescale on which it fragments. Moreover, there is a close relationship between the flux of matter determining the accumulation timescale ($\sim\!\rho\subF v\subF$), and the ram-pressure determining the density in the layer ($\sim\!\rho\subF v\subF^2$). The results derived by \citet{LarsonRB1985} (and others), which start with an already accumulated layer, are not applicable to this situation, because gravity is not well-mannered, it does not politely hold back until a layer is accumulated and then do its thing. From the moment a layer starts accumulating, gravity unrelentingly looks for opportunities to fragment the layer. As the layer accumulates, the growth time of the most unstable wavelength decreases monotonically, and non-linear fragmentation sets in as soon as there is a perturbed wavelength whose growth time is comparable with the elapsed time. The upshot is that the layer starts to fragment long before it becomes sufficiently massive and thick to be susceptible to the instability analysed by \citet{LarsonRB1985}. We therefore refer to the fragmentation mode analysed by Larson as the late fragmentation mode, and to that analysed here as the early fragmentation mode.

Specifically, non-linear fragmentation of an accumulating layer is initiated by corrugation waves condensing into filaments, and starts after the layer has been accumulating for a time $\;t_{_{\rm FRAG.EARLY}}\!\sim\!1/(G\rho\subF{\cal M})^{1/2}\;$ (see Eqn. \ref{EQN:tCRIT1}, and recall that ${\cal M}=v\subF/a\subL$). At this juncture, the layer thickness is only $\;2Z_{_{\rm EARLY}}\!\sim\!a\subL/(G\rho\subF{\cal M}^3)^{1/2}\;$ (see Eqn. \ref{EQN:Thickness}), but the fragmentation wavelength in the plane of the layer is $\;2Y_{_{\rm EARLY}}\!\sim\!a\subL/(G\rho\subF{\cal M})^{1/2}\;$ (see Eqn. \ref{EQN:2YCRIT}), so the proto-filaments have cross-sections with an aspect ratio of order $2Y_{_{\rm EARLY}}/2Z_{_{\rm EARLY}}\!\sim\!{\cal M}$ (see Eqn. \ref{EQN:Y2Z}). Thus, the higher the Mach Number, ${\cal M}$, the more markedly the initial motions assembling a filament are parallel to the plane of the layer, and the larger the ratio between the width of the final filament and the distance to the next filament.

In contrast, the fragmentation mode analysed by \citet{LarsonRB1985} can only develop once the layer is sufficiently thick to contain a standard Jeans wavelength, evaluated at the density in the layer, i.e. $\;2Z_{_{\rm LATE}}\!\sim\!a\subL/(G\rho\subF{\cal M}^2)^{1/2},\;$ and this is why we call it the late fragmentation mode. In the late fragmentation mode, the fastest growing wavelength parallel to the plane of the layer is somewhat larger than the thickness of the layer, and the fragmentation timescale is very short, $t_{_{\rm FRAG.LATE}}\!\sim\!1/(G\rho\subF{\cal M}^2)^{1/2},\;$ but it takes such a long time to accumulate a layer that is sufficiently thick for the late mode to develop, $t_{_{\rm ACCUM.LATE}}\!\sim\!1/(G\rho\subF)^{1/2},\;$ that the late mode will normally be pre-empted by the early mode.

\subsection{R{\'e}sum{\'e}}

We now explore in more detail the basic phenomenology of turbulent fragmentation, and how it might operate in practice. We start by developing a model for the formation and fragmentation of a cool shock-compressed layer (Section \ref{SEC:DYNAMICS}). Then we formulate the cooling rates delivered by molecules and dust, and show that molecules dominate post-shock cooling just inside the boundaries of the layer, but dust dominates the cooling of condensing low-mass prestellar cores (Section \ref{SEC:THERMO}). It should be noted that these results regarding the dominant cooling channels do not depend on the details of the model for formation and fragmentation of a shock compressed layer -- except in as much as (i) there is shock compression, and (ii) cores condense out of gas at the post-shock density. We also re-iterate that, between (a) the shock compression and post-shock cooling by molecules at the layer boundary, and (c) subsequent core collapse with dust cooling, there is an extended period (b) during which gas collects into filaments and then into cores at approximately constant density.

\section{The formation and fragmentation of shock-compressed layers}\label{SEC:DYNAMICS}

In this section we consider the flows that generate shock-compressed layers, and the structure of the resulting layers. The key result is that there is an {\it early phase}, $t\!<\!t_{_{\rm EARLY}}$, during which the layer is held together mainly by the ram pressure of the inflowing gas, and the density in the layer is approximately uniform. Then, we consider the flows that fragment a shock-compressed layer. During the early phase, the layer becomes unstable against gravitational fragmentation, and the principal mode of fragmentation is a corrugation wave that breaks the layer up into  quite widely spaced filaments, and thence into cores. We reiterate that we are here concerned with an accumulating layer, rather than a static, already accumulated layer \citep[cf.][]{LarsonRB1985,Miyaetal1987a,Miyaetal1987b}. In other words, matter continues to flow into the layer, and the layer {\it simultaneously} tries to fragment gravitationally into filaments. \citet{ElmeElme1978} do explicitly refer to the trade-off between the timescale for accumulation of a layer and the timescale on which it fragments, but do not explore the consequences. As the surface-density of the layer increases, the timescale on which the fastest growing corrugation wave develops decreases, and the layer breaks up into filaments once this timescale becomes less than the time for which the layer has been growing. At their inception, the separation between neighbouring filaments is much greater than the thickness of the layer, by a factor $\sim\!{\cal M}$. This section uses results from Appendices \ref{APP:gY} and \ref{APP:CYL}.

\subsection{Convergent flows}

Consider two anti-parallel flows of gas, each with uniform density $\rho\subF$, and (in the centre-of-mass frame) uniform velocities ${\bf v}\!=\!(0,0,\pm v\subF$), colliding head-on. Exactly matched, uniform-density, anti-parallel flows are evidently a crude idealisation, one of several introduced in this analysis in order to ensure simple equations that, notwithstanding, capture the essential elements of layer fragmentation. In real molecular clouds the flows are turbulent and chaotic, but a significant layer can only be formed where, in the rest-frame of the layer, two supersonic flows of comparable ram-pressure meet. Therefore the estimates we derive should be representative.\footnote{That is, unless pre-existing structures in the inflow are sufficiently large to determine the scale of fragmentation. In this case one might argue that the inflowing gas is already fragmented. The issue of how large such pre-existing structures need to be to do this is an interesting one. However, it can only be addressed numerically, and we plan to do this in a future paper. In the meantime, the existence of very large coherent filaments in the interstellar medium suggests that there must be circumstances under which pre-existing structure is either absent or quickly annealed by dissipation.} Where the flows meet, a dense layer forms, bounded by two accretion shocks at $z\!=\!\pm\,Z\subL(t)$. The gas-kinetic temperature in the bulk of the layer is $T\subL$, with corresponding isothermal sound-speed $a\subL$.

\subsection{Hydrostatic balance for a plane-parallel isothermal gas layer}\label{SEC:HB}

To analyse the structure of the layer, we first consider a plane-parallel isothermal layer in hydrostatic balance. The density at distance $z$ above, or below, the midplane is
\begin{eqnarray}\label{EQN:RHOZ}
\rho\subL(z)&=&\rho\subL(0)\;{\rm sech}^2\!\left(\!\frac{|z|}{z\subO}\!\right)
\end{eqnarray}
\citep{LedouxP1951}, where the scale-height
\begin{eqnarray}\label{EQN:zO}
z\subO&=&\frac{a\subL}{(2\pi G\rho\subL(0))^{1/2}}\,.
\end{eqnarray}

If the layer is truncated at $z=\pm Z\subL$, the column-density through the layer is
\begin{eqnarray}\label{EQN:SIGZ}\nonumber
\Sigma\subL&=&\int\limits_{z=-Z\subL}^{z=+Z\subL}\,\rho\subL(z)\,dz\\
&=&2\,\rho\subL(0)\,z\subO\,{\rm tanh}\!\left(\!\frac{Z\subL}{z\subO}\!\right)\,,
\end{eqnarray}
and it is still in equilibrium, provided that it is contained by an external pressure
\begin{eqnarray}\label{EQN:PEXT1}
P_{_{\rm EXT}}&=&\rho\subL(Z\subL)\,a\subL^2\,.
\end{eqnarray}

If we now substitute for ${\rm sech}(Z\subL/z\subO)$ from Eq. (\ref{EQN:RHOZ}), and for ${\rm tanh}(Z\subL/z\subO)$ from Eq. (\ref{EQN:SIGZ}), in ${\rm sech}^2\!=\!1-{\rm tanh}^2$, and for $z_{_{\rm O}}$ from Eq. (\ref{EQN:zO}), we obtain
\begin{eqnarray}\nonumber
\rho\subL(Z\subL)&=&\rho\subL(0)\,-\,\frac{\pi\, G\,\Sigma\subL^2}{2\,a\subL^2}\,,
\end{eqnarray}
or, with $\rho\subL(Z\subL)\!\rightarrow\!\rho_{_{\rm SURFACE}}$ and $\rho\subL(0)\!\rightarrow\!\rho_{_{\rm MIDPLANE}}$, 
\begin{eqnarray}\label{EQN:RHOMID1}
\rho_{_{\rm MIDPLANE}}&=&\rho_{_{\rm SURFACE}}\,+\,\frac{\pi\, G\,\Sigma_{_{\rm LAYER}}^2}{2\,a\subL^2}\,.
\end{eqnarray}
The second term on the righthand side of Eq. (\ref{EQN:RHOMID1}) measures the contribution that self-gravity makes to holding the layer together. If it is much larger than $\rho_{_{\rm SURFACE}}$, self-gravity is dominant and there needs to be a large pressure gradient in the $z$ direction to support the layer, so the midplane is significantly denser than the surface. Conversely, if the second term is much smaller than $\rho_{_{\rm SURFACE}}$, self-gravity makes only a small contribution to holding the layer together, and the density on the midplane is only slightly larger than the density at the surface.

From Eq. (\ref{EQN:PEXT1}),
\begin{eqnarray}\label{EQN:RHOSURF}
\rho_{_{\rm SURFACE}}&=&\frac{P_{_{\rm EXT}}}{a\subL^2}\,,
\end{eqnarray}
and Eq. (\ref{EQN:RHOMID1}) reduces to
\begin{eqnarray}\label{EQN:RHOMID2}
\rho_{_{\rm MIDPLANE}}&=&\frac{P_{_{\rm EXT}}}{a\subL^2}\,+\,\frac{\pi\,G\,\Sigma_{_{\rm LAYER}}^2}{2\,a\subL^2}\,.
\end{eqnarray}
Therefore, as long as 
\begin{eqnarray}\label{EQN:EARLY1}
\Sigma_{_{\rm LAYER}}&\ll&\left(\frac{2P_{_{\rm EXT}}}{\pi G}\right)^{\!1/2}\,,
\end{eqnarray}
self-gravity plays only a small role in holding the layer together in the $z$-direction; the density in the layer is then approximately uniform, $\rho_{_{\rm MIDPLANE}}\!\sim\!\rho_{_{\rm SURFACE}}$, and is mainly determined by the external pressure (Eq. \ref{EQN:RHOSURF}).

\subsection{Layers created by colliding flows}

For a layer created by colliding supersonic flows, $v\subF\gg a\subL$, so the external pressure is dominated by the ram-pressure of the inflowing gas,
\begin{eqnarray}\label{EQN:PEXT2}
P_{_{\rm EXT}}&\sim&\rho\subF v\subF^2\,,
\end{eqnarray}
and the density at the surface of the layer (Eq. \ref{EQN:RHOSURF}) is
\begin{eqnarray}
\rho_{_{\rm SURFACE}}&\sim&\frac{\rho\subF v\subF^2}{a\subL^2}\,.
\end{eqnarray}

The surface-density of the layer is
\begin{eqnarray}\label{EQN:Sigt}
\Sigma_{_{\rm LAYER}}(t)&\sim&2\rho\subF v\subF t\,,
\end{eqnarray}
so the condition for the layer to have approximately uniform density (Eq. \ref{EQN:EARLY1}) reduces to
\begin{eqnarray}
t&<&t_{_{\rm EARLY}}\;\,\sim\;\,(2\pi G\rho\subF)^{-1/2}\,,
\end{eqnarray}
where $t_{_{\rm EARLY}}$ is essentially the freefall time in the pre-shock gas. We refer to the phase during which the growing layer has approximately uniform density as the {\it early phase}, and the analysis we present below is concerned largely with this early phase. This is because, as long as the gas is approximately isothermal, the layer fragments gravitationally during the early phase, and hence the layer never makes it to the late phase (where self-gravity would become important in holding the layer together in the $z$-direction). This apparently paradoxical statement reflects the fact that the self-gravity which drives fragmentation of the layer is predominantly acting parallel to the plane of the layer (perpendicular to $\hat{z}$), and involves relatively large lengths ($2Y_{_{\rm FRAG.ANISOT}}$, see Eqn. \ref{EQN:2YCRIT}) and hence relatively large column-densities of matter. In contrast, the self-gravity that contributes rather weakly to holding the layer together acts perpendicular to the layer (parallel to $\hat{z}$) and involves relatively small lengths ($2Z_{_{\rm FRAG.ANISOT}}$ and Eqn. \ref{EQN:Thickness}) and hence relatively small column-densities. We expand upon this issue in Section \ref{SEC:DOMWL}.

Since the density is approximately uniform during the early phase, the half-thickness of the layer is
\begin{eqnarray}
Z_{_{\rm LAYER}}(t)&\sim&\frac{\Sigma_{_{\rm LAYER}}(t)}{2\rho_{_{\rm SURFACE}}}\;\,\sim\;\,\frac{a\subL^2 t}{v\subF}\,.
\end{eqnarray}
The sound travel time between the surface of the layer and its midplane,
\begin{eqnarray}
t_{_{\rm ST}}(t) \sim \frac{Z_{_{\rm LAYER}}(t)}{a\subL} \sim \frac{a\subL t}{v\subF}\,,
\end{eqnarray}
is much less than the elapsed time, $t$ (recall that $v\subF\gg a\subL$). Consequently, the accumulating layer has plenty of time to relax towards hydrostatic balance {\it in the $z$-direction}. The assumption that -- prior to the development of the corrugation instability -- the gas in the layer is close to hydrostatic equilibrium is therefore reasonable. This also suggests that turbulence will be significantly damped in the layer. Hereafter we ignore density variation in the layer, and put
\begin{eqnarray}\label{EQN:RHOS}
\rho_{_{\rm MIDPLANE}}\,\sim\, \rho_{_{\rm SURFACE}}\,\sim\,\frac{\rho\subF v\subF^2}{a\subL^2}\,\equiv\,\rho\subL.
\end{eqnarray}

\subsection{The geometry of a proto-filament}\label{SEC:FIL1}

Consider a plank-shaped proto-filament cut from a layer of approximately uniform density. Since we are in the early phase, the density in the layer is set by the ram pressure (Eq. \ref{EQN:RHOS}). Without loss of generality, we assume that the proto-filament has infinite length in the $x$-direction (which is in the plane of the layer). It is plank-shaped in the sense that it has a rectangular cross-section (see Fig. 1a) and its dimension in the $y$-direction (which is also in the plane of the layer, $\,-Y\!<\!y\!<\!Y$) is significantly larger than its dimension in the $z$-direction (which is the thickness of the layer, $\,-Z\subL\!<\!z\!<\!Z\subL$). In other words, the wavelength of the corrugation instability that assembles a proto-filament is significantly greater than the thickness of the layer, $\,2Y\!\gg\!2Z\subL$. Because the density is approximately constant at the value given by Eq. (\ref{EQN:RHOS}), the motions amplifying a proto-filament start off parallel to the $y$-axis, but then diverge as they approach the symmetry axis of the filament (see Fig. 1b).

\begin{figure}
\begin{center}
\includegraphics[width=1.25\columnwidth,angle=270]{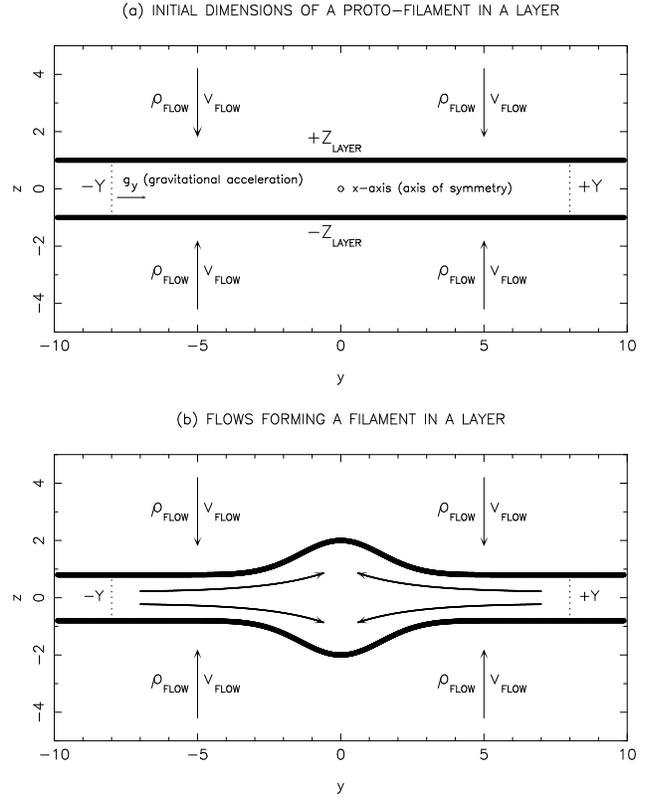}
\caption{(a) The initial conditions for a proto-filament to condense out of a shock compressed layer. The layer is between $z=\pm Z\subL$ (where the $z$-axis is vertical), and gas with density $\rho\subF$ flows into the layer at velocity $v\subF$, parallel and anti-parallel to the $z$-axis. The $x$-axis is perpendicular to the image plane, and the proto-filament is infinite in this direction. The proto-filament is initially between $y=\pm Y$, and $Y\gg Z\subL$. $\;g_y$ is the gravitational acceleration, in the $y$-direction which acts to amplify the proto-filament. (b) The flows assembling a filament are mainly in the $y$-direction, but because the density in the layer is fixed by the ram-pressure of the inflowing gas, the layer has to bulge to accommodate the gas flowing into the filament.}\label{FIG:Geometry}
\end{center}
\end{figure}

\subsection{Condensation of a proto-filament}\label{SEC:CONDPF}

The equation of motion for the half-width of the proto-filament is, approximately,
\begin{eqnarray}\nonumber
{\ddot Y}&\sim&-\,8\,G\,\Sigma_{_{\rm LAYER}}\,+\,\frac{a\subL^2}{Y}\,;
\end{eqnarray}
the first term on the righthand side represents self-gravity acting parallel to the $y$-axis (i.e. $g_{_y}(+Y)\!=\!-g_{_y}(-Y)$ from Eq. \ref{EQN:gy2}), and the second term represents pressure.

Condensation of the proto-filament requires $\ddot{Y}<0$, i.e.
\begin{eqnarray}
Y&\ga&\frac{a\subL^2}{8G\Sigma\subL}\,,
\end{eqnarray}
and for proto-filaments satisfying this constraint, the condensation timescale is given by
\begin{eqnarray}\nonumber
t_{_{\rm COND.EARLY}}(Y)\!&\!\sim\!&\!\left(\frac{2Y}{-{\ddot Y}}\right)^{1/2}\\
\!&\!\sim\!&\!\left(\frac{4G\Sigma_{_{\rm LAYER}}}{Y}-\frac{a\subL^2}{2Y^2}\right)^{\!-1/2}.
\end{eqnarray}

Therefore, at time $t$, the fastest growing proto-filament has initial wavelength
\begin{eqnarray}\label{EQN:2Y_FASTEST}
2Y_{_{\rm FASTEST.EARLY}}(t)&\sim&\frac{a\subL^2}{2G\Sigma_{_{\rm LAYER}}(t)}\,,
\end{eqnarray}
and condensation timescale
\begin{eqnarray}\label{EQN:t_FASTEST}
t_{_{\rm FASTEST.EARLY}}(t)&\sim&\frac{a\subL}{2^{3/2}G\Sigma_{_{\rm LAYER}}(t)}\,.
\end{eqnarray}

\subsection{The dominant wavelength for fragmentation of an accumulating layer into filaments}\label{SEC:DOMWL}

As the surface-density of the layer, $\Sigma_{_{\rm LAYER}}(t)$, increases (see Eq. \ref{EQN:Sigt}), the fastest growing wavelength, $2Y_{_{\rm FASTEST.EARLY}}(t)$ (Eq. \ref{EQN:2Y_FASTEST}), and its condensation timescale, $t_{_{\rm FASTEST.EARLY}}(t)$ (Eq. \ref{EQN:t_FASTEST}), both decrease monotonically. Combining Eqs. (\ref{EQN:t_FASTEST}) and (\ref{EQN:Sigt}), we have
\begin{eqnarray}\nonumber
t_{_{\rm FASTEST.EARLY}}(t)&\sim&\frac{a\subL}{2^{5/2}G\rho\subF v\subF t},
\end{eqnarray}
and non-linear fragmentation starts around the time when this timescale becomes shorter than the elapsed time, $t$, i.e. at
\begin{eqnarray}\label{EQN:tCRIT1}
t_{_{\rm FRAG.EARLY}}&\sim&\left(\frac{a\subL}{2^{5/2}\,G\,\rho\subF\,v\subF}\right)^{1/2}\,.
\end{eqnarray}

The critical wavelength for fragmentation is then
\begin{eqnarray}\label{EQN:2YCRIT}
2Y_{_{\rm FRAG.EARLY}}&\sim&\left(\frac{a\subL^3}{2^{3/2}\,G\,\rho\subF\,v\subF}\right)^{1/2}\,,
\end{eqnarray}
and when non-linear fragmentation starts, the ratio of the critical wavelength to the thickness of the layer is
\begin{eqnarray}\label{EQN:Y2Z}
\frac{2Y_{_{\rm FRAG.EARLY}}}{2Z\subL(t_{_{\rm FRAG.EARLY}})}&\sim&\frac{v\subF}{2^{1/2}\,a\subL}
\end{eqnarray}
(as assumed in Appendix \ref{APP:gY}). In other words, at its inception, the cross-section of a critical proto-filament  (looking along its symmetry axis, the $x$-axis) has a large aspect ratio; it is a plank rather than a pole. Consequently, the motions assembling the filament are initially mainly parallel to the $y$-axis, and hence parallel to the plane of the layer. However, since during the early phase the density of the shocked gas is still approximately uniform and controlled by the ram pressure of the inflowing gas (see Eq. \ref{EQN:RHOS}), the gas flowing across the layer and into the filament also has to diverge in directions perpendicular to the plane of the layer (the $z$-direction; see Fig. 1b) so as to maintain this density.

We note that both the critical wavelength for early fragmentation of an accumulating layer into filaments (Eq. \ref{EQN:2YCRIT}), and the timescale on which such a layer fragments into filaments (Eq. \ref{EQN:tCRIT1}), are larger, by a factor $\sim\!(v\subF/a\subL)^{1/2}$, than the corresponding quantities for {the late fragmentation of an already accumulated layer, as analysed by, for example, \citet{LarsonRB1985},}
\begin{eqnarray}\nonumber
2Z_{_{\rm FRAG.LATE}}\!&\!\sim\!&\!\frac{a\subL}{(G\rho\subL)^{1/2}}\;\sim\;\frac{a\subL^2}{(G\rho\subF)^{1/2}v\subF}\,,\\\label{EQN:lambdaJ3D}\\\nonumber
t_{_{\rm FRAG.LATE}}\!&\!\sim\!&\!\frac{1}{(G\rho\subL)^{1/2}}\;\sim\;\frac{a\subL}{(G\rho\subF)^{1/2}v\subF}\,.\\\label{EQN:tJ3D}
\end{eqnarray}
Nonetheless, it is the early fragmentation mode that actually occurs, because the layer becomes massive and thick enough to sustain the early fragmentation mode, i.e.
\begin{eqnarray}\label{EQN:Thickness}
2Z\subL(t_{_{\rm FRAG.EARLY}})&\sim &\left(\frac{a\subL^5}{G\rho\subF v\subF^3}\right)^{1/2}\,,
\end{eqnarray}
long before it becomes massive and thick enough (i.e. $2Z_{_{\rm FRAG.LATE}}$, see Eq. \ref{EQN:lambdaJ3D}) to sustain the late fragmentation mode. The critical issue is that the time it takes to collect a layer that is massive and thick enough to sustain the late fragmentation mode is much {\it longer} (again by a factor ${\cal M}^{1/2}\!\sim\!(v\subF/a\subL)^{1/2}$) than the time it takes to collect a layer that is massive and thick enough to support the early fragmentation mode.

\subsection{Filament orientation}

When an accumulating layer fragments into filaments, the orientation of the filaments is random, unless it is influenced by some anisotropic force. For example, if there is a uniform large-scale magnetic field, ${\bf B}$, or a large-scale angular velocity, ${\boldsymbol\omega}$ (say, due to shear in the collision producing the layer), with a component in the plane of the layer, motions perpendicular to ${\bf B}$ or ${\boldsymbol\omega}$ are inhibited, and so the first filaments tend to be aligned perpendicular to ${\bf B}$ or ${\boldsymbol\omega}$, and hence parallel to one another. There are perturbations in all directions, it's just that the ones involving motions perpendicular to ${\bf B}$ or ${\boldsymbol\omega}$ tend to grow more slowly, and hence to become non-linear later. For simplicity, we proceed on the assumption that there is no preferred filament orientation, and therefore filaments form with random orientations and frequently intersect. Cores then tend to form with essentially the same spacing along filaments as between filaments, i.e. $\sim 2Y_{_{\rm FRAG.EARLY}}$ (see Eq. \ref{EQN:2YCRIT}).

\subsection{A possible schema for the evolution of a star-forming filament}\label{SEC:TOY1}

We assume that it takes from $t=0$ to $t\!\sim\!t_{_{\rm FRAG.EARLY}}$ for non-linear fragmentation of a layer to start, and from $\sim\!t_{_{\rm FRAG.EARLY}}$ to $\sim\!2t_{_{\rm FRAG.EARLY}}$ to establish the lateral flows within the layer that build a filament. Thereafter, until the supply ceases, matter flows across the layer and into the filament at roughly the same rate that it flows onto the part of the layer that is feeding the filament (see Fig. 1b), so that during this period the line-density of the filament grows at a rate 
\begin{eqnarray}\nonumber
{\dot \mu}_{_{\rm IN}}&\sim&4Y_{_{\rm FRAG.EARLY}}\rho\subF v\subF\\\label{EQN:mudot}
&\sim&\left(\!\frac{2^{1/2}a\subL^3\rho\subF v\subF}{G}\!\right)^{\!1/2}.
\end{eqnarray}

By the time the line-density of a filament reaches $\sim\!\mu_{_{\rm MAX}}/2\!=\!a\subL^2/G$ (see Eq. \ref{EQN:muMAX}) at $\sim\!3t_{_{\rm FRAG.EARLY}}$, flows along the filament have started to develop, delivering matter towards proto-cores. These proto-cores are typically separated by $\sim 2Y_{_{\rm FRAG.EARLY}}$, since they are usually located where the original filament is crossed by a second, approximately orthogonal filament that developed a bit more slowly, by virtue of being seeded by a smaller perturbation. After $\sim\!4t_{_{\rm FRAG.EARLY}}$, and until the supply ceases, matter flows along filaments into cores at roughly the same rate as it flows into the section of filament that is feeding the core, so the mass of a core grows at a rate
\begin{eqnarray}
{\dot M}&\sim&2Y_{_{\rm FRAG.EARLY}}{\dot \mu}_{_{\rm IN}}\;\,\sim\;\,\frac{a\subL^3}{2^{1/2}G}\,.
\end{eqnarray}
The line-density of the parent filament is approximately constant at
\begin{eqnarray}
\mu_{_{\rm MAX}}&\sim&\frac{3a\subL^2}{2G}\;\,\sim\;\,13\,{\rm M}_{_\odot}\,{\rm pc}^{-1}\,\left(\frac{T\subL}{10\,{\rm K}}\right)\,,
\end{eqnarray}
but with a turn-over time of $\sim\!2t_{_{\rm FRAG.EARLY}}$ due to matter flowing into the filament from the sides, and then along the filament into cores.

Once the filament is established ($t\ga 3t_{_{\rm FRAG.EARLY}}$), and until the supply of matter ceases, its radius is given approximately by $\pi R_{_{\rm FIL}}^2\rho\subL\sim \mu_{_{\rm MAX}}$, hence its diameter is
\begin{eqnarray}\nonumber
2R_{_{\rm FIL}}&\sim&\left(\!\frac{6a\subL^2}{\pi G\rho\subL}\!\right)^{1/2}\\\nonumber
&\la&\left(\frac{6}{\pi G P_{_{\rm CRIT}}}\right)^{1/2}\,a\subL^2\\\label{EQ:R_FIL_1}
&\la&0.16\,{\rm pc}\,\left(\frac{T\subL}{10\,{\rm K}}\right)\,.
\end{eqnarray}

The ratio of the filament's diameter to the separation between neighbouring filaments (equivalently, the fraction by which a filament contracts in the $y$-direction as it converts from a plank to a pole) is
\begin{eqnarray}
\frac{2R_{_{\rm FIL}}}{2Y_{_{\rm FRAG.EARLY}}}&\sim&\left(\frac{2^{9/2} a\subL}{\pi v\subF}\right)^{1/2}\,.
\end{eqnarray}
If this ratio can be measured observationally, it gives a constraint on the Mach Number of the accretion shock at the boundary of the birth layer -- specifically, in view of projection, a lower limit on the Mach Number.

The core will become gravitationally unstable, as soon as its mass exceeds the standard Jeans mass (Eq. \ref{EQN:MBEM}) corresponding to the density $\rho\subL$ (set by the ram pressure of the gas flowing into the layer, Eq. \ref{EQN:RHOS}),  i.e.
\begin{eqnarray}\nonumber
M&\sim&\frac{1.87\,a\subL^4}{\left(G^3\,P_{_{\rm CRIT}}\right)^{1/2}}\\
&\sim&1\,{\rm M}_{_\odot}\,\left(\frac{T\subL}{10\,{\rm K}}\right)^2
\end{eqnarray}
Once the flows across the layer and into the filament have been established, the time to form a gravitationally unstable core is 
\begin{eqnarray}
\Delta t&\sim&\frac{M}{\dot{M}}\;\,\sim\;\,\frac{2.65\,a\subL}{(G\rho\subF)^{1/2}\,v\subF}.
\end{eqnarray}
This is thus much shorter than the time it takes to set up the flows, $t_{_{\rm FRAG.EARLY}}$ (Eqn. \ref{EQN:tCRIT1}), by a factor ${\cal M}^{-1/2}\!\sim\! (a\subL/v\subF)^{1/2}$. 

As more matter flows into a core, it can either accrete onto existing protostars or form additional ones. This issue lies outside the scope of the present paper, but presumably it will be influenced by the specific angular momentum of the inflowing material, and feedback from the existing protostars. The simulations of \citet{Girietal2012a} suggest that the additional matter forms new protostars.

\subsection{Edge effects and other caveats}

This paper is limited to consideration of two-dimensionally infinite shock-compressed layers produced by colliding flows, and one-dimensionally infinite filaments condensing out of them. A number of authors \citep[e.g.][]{BurkHart2004,PonAetal2012} have pointed out that the strongly focused gravitational field at the edge (or end) of a finite, truncated layer (or filament) can lead to the growth of large condensations at these positions, and this effect has been reproduced in numerical simulations \citep[e.g.][]
{NelsPapa1993,HartBurk2007,ClarWhit2015,SeifWalc2015}.

In the context of shock compressed layers produced by colliding streams (as considered in this paper), it is questionable whether this edge fragmentation mode is relevant. The two flows creating such a layer are unlikely to be so well matched in their lateral extents that they form a layer with a well defined boundary. Rather, the edge of the resulting layer will usually be poorly defined, {\it and} strongly sheared as material from the more extended flow continues past the less extended one. Perhaps a more germane issue is the overall lateral contraction of a finite shock-compressed layer produced by colliding streams, i.e. `global collapse' in the terminology of \citet{PonAetal2011}. In the simulations of colliding clouds discussed by \citet{Balfetal2015}, the resulting shock-compressed layer fragments into a network of filaments. If the collision is relatively fast, the fragmentation is essentially the same as for a patch on an infinite layer. However, if the collision is relatively slow, it takes longer for the layer to reach a sufficient column-density to fragment, and, by the time it does so, the layer is contracting laterally; consequently the filaments are dragged towards the centre of mass, and form a hub-and-spoke system, feeding newly formed stars and residual material into a central star cluster.

In the context of filament formation and fragmentation, it is ambiguous what role the end mode plays in nature. There are a few cases where very massive very extended filamentary infrared dark clouds (IRDCs) appear to be dominated by end clumps: the Nessie nebulae, which is over $\sim\!80\,{\rm pc}$ long \citep{Jacketal2010}, has giant H{\sc ii} regions at both ends, implying a significant outburst of star formation at these two locations in the recent past; the two most massive clumps in the main filament in NGC6334, which is $\sim\!12\,{\rm pc}$ long, are located at its ends \citep{Zernetal2013}; the Musca filament, which is $\sim\!6\,{\rm pc}$ long, appears to have fragmented more rapidly at its ends \citep{Kainetal2015}; the filamentary IRDC18223, which is $\sim\!4\,{\rm pc}$ long \citep{Beutetal2015}, is dominated by two end-clumps. However, there are many smaller filaments, in nearby low-mass star formation regions, where the ends do not appear to have played an important role in fragmentation, and the prestellar cores are distributed more-or-less regularly along the filament \citep[e.g. in Taurus,][]{HacaTafa2011,Hacaetal2013}. \citet{PonAetal2011} stress that in a filament the timescale for local collapse (producing periodically-spaced cores) can be shorter than the timescale for global collapse.

\citet{HeitschF2013a,HeitschF2013b} has explored the consequences of cylindrically symmetric accretion onto filaments, the resulting ram pressure (amplified by freefall acceleration), and the possible influence of turbulence in filaments (generated by such accretion flows). In some regards, this work relates to the analysis presented here for the growth and fragmentation of a shock-compressed layer, but there are also significant differences. For a shock-compressed layer, as analysed here, the geometry is simple, and there is a trivial relation between the rate of growth of the layer ($\sim\!\rho\subF v\subF$) and the ram pressure that it delivers at the accretion shock ($\sim\!\rho\subF v\subF^2$). This makes it very straightforward to evaluate the competition between the timescale for growth of the layer, and the timescale for its fragmentation. Our analysis also suggests that turbulence will probably be significantly damped in the layer. For the filaments analysed by Heitsch, the converging geometry of the flow makes the analysis much more complicated, and -- as Heitsch emphasises -- the relationship between the different quantities characterising the configuration is quite complicated. Clarke et al. (in prep.) show that the spacing of condensations along a filament allows one to constrain how long a filament has been accreting, and at what rate, because an accreting filament undergoes gravo-acoustic oscillations, and the fastest growing condensations along the filament are the ones that become unstable in phase with these oscillations. This result can be applied to the filaments condensing out of shock compressed layers in the schema outlined in Section \ref{SEC:TOY1}. However, these filaments are very different from the ones analysed by \citet{HeitschF2013a,HeitschF2013b}, in that they grow from material flowing in anisotropically, mainly parallel to the plane of the mother-layer, at approximately constant density (see Fig. \ref{FIG:Geometry}) -- rather than from a cylindrically isotropic convergent flow.

\section{The cooling of gas in shock-compressed layers and condensing prestellar cores}\label{SEC:THERMO}

In the turbulent fragmentation picture of star formation that we have sketched above, there are two critical cooling requirements. Initially, where gas streams collide to produce shock-compressed layers, the layers will only fragment gravitationally (in the manner described in Sections \ref{SEC:FIL1} through \ref{SEC:DOMWL}), if the post-shock gas that has just joined the layer cools radiatively to a low temperature, on -- or preferably faster than -- the fragmentation timescale. Later on, the gravitationally unstable cores, resulting from layer fragmentation and the flow of shocked gas into and along filaments, will only condense into stars if they can cool radiatively, on a dynamical -- i.e. approximately freefall -- timescale. (In between, cooling is unimportant because the flows of gas into and along filaments that create cores occur at approximately constant density.)

To evaluate these conditions we need expressions for the basic thermal properties of star-forming gas, and in particular for the radiative cooling rates. We assemble these expressions here, using results from Appendices \ref{APP:CO} (CO line cooling) and \ref{APP:DUST} (dust continuum cooling). We use these cooling rates to show that molecular-line cooling is in general more effective than dust cooling for the gas flowing into the accretion shock at the boundary of a shock-compressed layer, and can easily cool the gas in the layer down to $T\subL\simeq 10\,{\rm K}$, fast enough for the layer to then fragment. Conversely, we find that in a collapsing prestellar core the capacity of line cooling is severely limited, by radiative trapping (i.e. high optical depth) and/or freeze-out, and here dust cooling dominates. It is this switch between line cooling in the post-shock region at the boundary of the layer, and dust cooling in a collapsing core, that defines the sweet spot for star formation, and hence the ram pressure threshold defined in Section \ref{SEC:BASICS}.

\subsection{Basic thermal properties of star-forming matter}\label{SEC:BASIC}

We are concerned here with star forming gas at number-densities in the range $100\,{\rm cm}^{-3}\la n_{_{{\rm H}_2}}\la 10^8\,{\rm cm}^{-3}$ (equivalently, mass-densities in the range $5\times 10^{-22}\,{\rm g}\,{\rm cm}^{-3}\la\rho\la 5\times 10^{-16}\,{\rm g}\,{\rm cm}^{-3}$), and temperatures in the range $10\,{\rm K}\la T\la 100\,{\rm K}$ (equivalently, isothermal sound speeds in the range $0.2\,{\rm km}\,{\rm s}^{-1}\la a\la 0.6\,{\rm km}\,{\rm s}^{-1}$). The metallicity, $Z$, abundance of dust by mass, $Z\subD$, mean mass per hydrogen molecule, ${\bar m}_{_{{\rm H}_2}}$, and mean gas-particle mass, ${\bar m}$, have been defined in Section \ref{SEC:INTRO}.

We assume that the dust can be represented by a single spherical grain type having radius $r\subD\sim10^{-5}\,{\rm cm}$, and internal density $\rho\subD\sim3\,{\rm g}\,{\rm cm}^{-3}$. It follows that the mass of a representative grain is $m\subD=4\pi r\subD^3\rho\subD/3$, the geometric cross-section of a representative grain is $\sigma\subD=\pi r\subD^2$, the number-density of grains is $n\subD=\rho Z\subD/m\subD$, and hence the product of the number-density of grains and the geometric cross-section of a grain is given by
\begin{eqnarray}\label{EQN:kappaD1}
n\subD\sigma\subD\!\!&\!\!=\!\!&\!\!\rho\,\kappa\subD\,,\\\label{EQN:kappaD2}
\kappa\subD\!\!&\!\!=\!\!&\!\!\frac{3Z\subD}{4r\subD \rho\subD}\,\sim\,2.35\!\times\! 10^2\,{\rm cm}^2\,{\rm g}^{-1}.
\end{eqnarray}

We assume that the absorption efficiency of a grain at far-infrared and submillimetre wavelengths can be approximated by
\begin{eqnarray}\label{EQN:QO}
Q(\lambda)&\sim&Q\subO\,\left(\!\frac{\lambda}{\rm cm}\!\right)^{\!-2}\,,
\end{eqnarray}
where $Q\subO\sim 2.5\times 10^{-7}$ (corresponding, for example, to a mass-opacity coefficient $\kappa_{_{250\mu{\rm m}}}\!\sim\!0.1\,{\rm cm}^2\,{\rm g}^{-1}$ for dust plus gas at $250\,\mu{\rm m}$). It follows that the Planck-mean emission efficiency of a grain at temperature $T_{_{\rm DUST}}$ is
\begin{eqnarray}\label{EQN:QPlanck}
{\bar Q}_{_{\rm PLANCK}}(T\subD)&\sim&{\bar Q}\subO\,\left(\frac{T\subD}{\rm K}\right)^{\!2}\,,
\end{eqnarray}
where ${\bar Q}\subO\sim 2.8\times 10^{-6}$.

In the range of density and temperature defined above, hydrogen molecules are effectively monatomic (because it is too cold to excite their rotational levels significantly). The only other abundant species is atomic helium, so the adiabatic exponent is $\gamma\simeq 5/3$, and the density of thermal energy is 
\begin{eqnarray}
u_{_{\rm THERM}}&\sim&n_{_{\rm TOT}}\frac{3k_{_{\rm B}}T}{2}\;\,=\;\;\frac{3\rho a^2}{2}\,.
\end{eqnarray}

\subsection{Cooling Rates}

As discussed in Appendix \ref{APP:CO}, the line cooling rate per unit volume is represented by the contribution from $^{12}$C$^{16}$O, which can be approximated with
\begin{eqnarray}\label{EQN:CO_TOT1}
\Lambda_{_{\rm CO.TOT}}&\sim&\left\{\Lambda_{_{\rm CO.LOW}}^{-1}+\Lambda_{_{\rm CO.HIGH}}^{-1}\right\}^{-1},
\end{eqnarray}
where 
\begin{eqnarray}\label{EQN:CO_LOW1}
\Lambda_{_{\rm CO.LOW}}&\sim&\left[{\rm Eq\,D4}\right]_{_{\rm CO.LOW}}\,\rho^2\,a^3
\end{eqnarray}
is the rate in the limit of low volume- and column-density (sub-thermal excitation, optically thin), and 
\begin{eqnarray}\label{EQN:CO_HIGH1}
\Lambda_{_{\rm CO.HIGH}}\!&\!\sim\!&\!\left[{\rm Eq\,D6}\right]_{_{\rm CO.HIGH}}a^8\frac{\Delta v}{L}
\end{eqnarray}
is the rate in the limit of high volume- and column-density (thermalised excitation, optically thick). The term $\Delta v/L$ is equivalent to $|dv/dz|$ in the Large Velocity Gradient (LVG) approximation.

At the lowest volume- and column-densities it is likely that line cooling is mainly due to O$^{\rm o}$, C$^{\rm o}$ and C$^+$, rather than CO, but our main concern here is the high-density regime. Here, CO is the dominant line coolant, but line cooling is becoming increasingly optically thick, {\it and} the gas-phase CO abundance is decreasing due to freeze-out onto dust. On both counts, the CO line cooling capacity of the gas is becoming increasingly limited, and the same considerations apply to other molecules, including the isotopomers of $^{12}$C$^{16}$O.

As discussed in Appendix \ref{APP:DUST}, the dust-cooling rate per unit volume can be approximated with
\begin{eqnarray}\label{EQN:GAS2DUST}
\Lambda\subD&\sim&\left[{\rm Eq\,E1}\right]_{_{\rm G2D}}\,\rho^2\,a^3\,.
\end{eqnarray}

\subsection{Jump conditions across a shock}

Consider a steady stream of gas with density $\rho\subF$, isothermal sound speed $a\subF$, and velocity ${\bf v}\!=\!(0,0,v\subF)$ flowing into a stationary, infinitesimally thin, planar shock-front fixed at $z\!=\!0$. For simplicity, we assume that the gas is and remains molecular, with adiabatic exponent $\gamma\!\simeq\!5/3$, and that $v\subF \gg a\subF$. Hence the gas emerges from the shock front with density, $\rho\subS$, velocity, $v\subS$, and isothermal sound-speed, $a\subS$, given by
\begin{eqnarray}\label{EQN:rhoS}
\frac{\rho\subS}{\rho\subF}&\simeq&\frac{8v\subF^2}{2v\subF^2+10a\subF^2}\;\,\sim\;\,4\,;
\end{eqnarray}
\begin{eqnarray}\label{EQN:vS}
\frac{v\subS}{v\subF}&\simeq&\frac{2v\subF^2+10a\subF^2}{8v\subF^2}\;\,\sim\;\,\frac{1}{4}\,;
\end{eqnarray}
and
\begin{eqnarray}\nonumber
a\subS^2&\simeq&\frac{\{2v\subF^2+10a\subF^2\}\{6v\subF^2-2a\subF^2\}}{64v\subF^2}\\\label{EQN:aS}
&\sim&\frac{3v\subF^2}{16}\,.\hspace{0.9cm}\\\nonumber
\end{eqnarray}
In Eqs. (\ref{EQN:rhoS}) through (\ref{EQN:aS}), the second expression on the righthand side is the approximate limiting form that obtains when $v\subF\!\gg\!a\subF$. Note that, for consistency, we continue to define the variable $a$ as the {\it isothermal} sound-speed, i.e. $a\!=\!(P/\rho)^{1/2}\!=\!(k_{_{\rm B}}T/\bar{m})^{1/2}$.

\subsection{The post-shock cooling zone}

In the post-shock radiative cooling zone, conservation of mass requires
\begin{eqnarray}\nonumber
\frac{d}{dz}\left\{\rho v \right\}&=&0\,,\\\label{EQN:MASSCON1}
\rho(z)&=&\frac{\rho\subF v\subF}{v(z)}\,;
\end{eqnarray}
conservation of momentum requires
\begin{eqnarray}\nonumber
\frac{d}{dz}\left\{\rho (v^2+a^2)\right\}\!&\!=\!&\!0\,,\\\label{EQN:MOMCON1}
a^2(z)\!&\!=\!&\!\left(\!v\subF\!+\!\frac{a^2\subF}{v\subF}\!\right)v(z)\!-\!v^2(z);
\end{eqnarray}
and conservation of energy requires
\begin{eqnarray}\nonumber
\frac{d}{dz}\left\{\rho v\left(\frac{v^2}{2}+\frac{\gamma a^2}{(\gamma-1)}\right)\right\}&=&-\,\Lambda(\rho,a)\,,
\end{eqnarray}
\begin{eqnarray}\nonumber
\rho\subF v\subF\!\left\{\frac{5}{2}\!\left(\!v\subF\!+\!\frac{a^2\subF}{v\subF}\!\right)\!-\!4v\right\}\!\frac{dv}{dz}\!&\!=\!&\!-\Lambda(\rho,a),\\\label{EQN:ENCON1}
\end{eqnarray}
where $\Lambda(\rho,a)$ is the radiative cooling rate per unit volume.

Since $v\subF\!\gg\! a\subF$, we have $v(z)\la v\subF /4$, and Eqs. (\ref{EQN:MOMCON1}) and (\ref{EQN:ENCON1}) approximate to
\begin{eqnarray}\label{EQN:MOMCON2}
a^2(z)&\sim&v\subF v(z)\,,\\\label{EQN:ENCON2}
\frac{5\,\rho\subF\,v\subF^2}{2}\frac{dv}{dz}&\sim&-\,\Lambda(\rho,a)\,.
\end{eqnarray}

The depth of the post-shock cooling region is then given by
\begin{eqnarray}\label{EQN:DeltaZ1}
\Delta Z^{^{\rm PSC}}&\sim&\int\limits_{v\sim v\subF/4}^{v\sim a\subL^2/v\subF}\,\frac{dv}{dv\!/\!dz}\,,
\end{eqnarray}
where the lower limit on the integral, $v\subF/4$, is the immediate post-shock velocity, and the upper limit, $a\subL^2/v\subF$, is the speed when the gas has cooled down to $T\subL=\bar{m}a\subL^2/k_{_{\rm B}}$. Similarly, the post-shock cooling time is given by
\begin{eqnarray}\label{EQN:Deltat1}
\Delta t^{^{\rm PSC}}&\sim&\int\limits_{v\sim v\subF/4}^{v\sim a\subL^2/v\subF}\,\frac{dv}{v\;dv\!/\!dz}\,.
\end{eqnarray}

\subsection{Post-shock CO cooling}\label{SEC:SHOCK_CO}

In the context of post-shock cooling, we substitute
\begin{eqnarray}
\frac{\Delta v}{L}&\rightarrow&\left|\frac{dv}{dz}\right|\;\,=\;\,-\,\frac{dv}{dz}\,,
\end{eqnarray}
and so Eq. (\ref{EQN:CO_HIGH1}) becomes
\begin{eqnarray}\label{EQN:CO_HIGH2}
\Lambda_{_{\rm CO.HIGH}}&\sim&[{\rm Eq\,D6}]_{_{\rm CO.HIGH}}\,a^8\,\left(\!-\frac{dv}{dz}\!\right)\,.
\end{eqnarray}

Next we substitute Eqs. (\ref{EQN:CO_LOW1}) and (\ref{EQN:CO_HIGH2}) into Eq. (\ref{EQN:CO_TOT1}), yielding
\begin{eqnarray}\nonumber
\Delta\Lambda_{_{^{12}{\rm C}^{16}{\rm O}}}^{^{\rm NET}}&\sim&\left\{\frac{1}{[{\rm Eq\,D4}]_{_{\rm CO.LOW}}\,\rho^2\,a^3}\right. \\\label{EQN:CO_TOT2}
&&\left.\;\;-\;\frac{1}{[{\rm Eq\,D6}]_{_{\rm CO.HIGH}}\,a^8\,dv\!/\!dz}\right\}^{-1}\,.
\end{eqnarray}
The first term in the braces on the righthand side of Eq. (\ref{EQN:CO_TOT2}) represents CO cooling in the low-density, optically-thin limit, and the second term represents CO cooling in the high-density, optically-thick limit (this term is preceded by a minus sign because $dv/dz$ is negative).

Finally, we substitute Eq. (\ref{EQN:CO_TOT2}) into Eq. (\ref{EQN:ENCON2}), to obtain
\begin{eqnarray}\nonumber
\left.\frac{dv}{dz}\right|_{_{\rm CO.TOT}}\!&\!\sim\!&\!-\,\frac{2\,[{\rm Eq\,D4}]_{_{\rm CO.LOW}}\,\rho\subF\, v\subF^{3/2}}{5\,v^{1/2}}\\\label{EQN:dvdz3}
&&\times\!\left\{1-\frac{5\,\rho\subF}{2\,[{\rm Eq\,D6}]_{_{\rm CO.HIGH}}\,v\subF^2\,v^4}\right\}\!.
\end{eqnarray}

In Eq. (\ref{EQN:dvdz3}), when the second term in the braces is negligible, the CO cooling is effectively in the low-density, optically thin limit, and the post-shock cooling time (Eq. \ref{EQN:Deltat1}) becomes
\begin{eqnarray}\label{EQN:Deltat_COLOW}
\Delta t^{^{\rm PSC}}_{_{\rm CO.LOW}}&\sim&\frac{5}{2\,[{\rm Eq\,D4}]_{_{\rm CO.LOW}}\,\rho\subF\,v\subF}\,.
\end{eqnarray}
If the layer is to fragment, $\Delta t^{^{\rm PSC}}_{_{\rm CO.LOW}}$ must be shorter than, or at least on the order of, the timescale on which the layer fragments, $t_{_{\rm FRAG.EARLY}}$ (see Eq. \ref{EQN:tCRIT1}), which reduces to
\begin{eqnarray}\label{EQN:THINSHOCK1}
n\subH \!\!&\!\ga\!&\!\!1.4\,{\rm cm}^{-3}\left(\frac{v\subF}{{\rm km}/{\rm s}}\right)^{-1}\left(\!\frac{T\subL}{10\,{\rm K}}\!\right)^{\!-1/2}\!.
\end{eqnarray}
This is evidently a very mild constraint, and is easily satisfied for the flows expected in turbulent molecular clouds.

As the gas in the post-shock region slows down, the second term in braces in Eq. (\ref{EQN:dvdz3}), $5\rho\subF/2[{\rm Eq\,D6}]_{_{\rm CO.HIGH}}v\subF^2v^4$, increases rapidly. If it becomes of order unity, CO line-cooling switches to the high-density, optically-thick limit, and can no longer cope. If this term is to remain less than unity until the gas decelerates to $v\sim a\subL^2/v\subF$, we have an additional constraint,
\begin{eqnarray}
n\subH \!\!&\!\la\!&\!\!3.4\!\times\! 10^3\,{\rm cm}^{-3}\!\left(\!\frac{v\subF}{{\rm km}/{\rm s}}\!\right)^{\!-2}\!\left(\!\frac{T\subL}{10\,{\rm K}}\!\right)^{\!4}\!.
\end{eqnarray}
This constraint is compatible with Eq. (\ref{EQN:THINSHOCK1}), but it restricts the pre-shock density to rather low values, particularly at high inflow velocities, $v\subF$. This is because we are requiring the gas to slow down to a very small post-shock velocity, $a\subL^2/v\subF$ (see Eq. \ref{EQN:Deltat1}). In reality, once the gas has slowed down to the local sound speed, the bandwidth available for line cooling probably derives mainly from residual trans-sonic turbulence. We therefore rework the above analysis, assuming that the critical requirement is for the post-shock gas to decelerate to $v\sim a\subL$ (rather than the smaller value, $v\sim a\subL^2/v\subF$) before the CO line cooling becomes optically thick. This gives
\begin{eqnarray}\label{EQN:THICKSHOCK1}
n\subH \!\!&\!\la\!&\!\!2.2\!\times\! 10^6\,{\rm cm}^{-3}\!\left(\frac{v\subF}{{\rm km}/{\rm s}}\right)^{\!2}\!\left(\!\frac{T\subL}{10\,{\rm K}}\!\right)^{\!2}\!,
\end{eqnarray}
which is easily satisfied for representative inflow velocities. 

We conclude that, even if we only take account of the contribution from $^{12}$C$^{16}$O, molecular-line cooling can bring the temperature of the post-shock gas back down to $T\subL\sim 10\,{\rm K}$, fast enough for the layer to fragment. If we allow for the effect of additional molecules, such as the isotopomers of CO, or if $T\subL$ is higher, as it might be in the vicinity of a luminous star cluster, the requirements (Eqs. \ref{EQN:THINSHOCK1} and \ref{EQN:THICKSHOCK1}) are even more easily satisfied.

\subsection{Post-shock dust cooling}\label{SEC:PS.DUST}

If we substitute Eq. (\ref{EQN:GAS2DUST}) into Eq. (\ref{EQN:ENCON2}), we obtain
\begin{eqnarray}
\left.\frac{dv}{dz}\right|_{_{\rm DUST}}&\sim&-\frac{2[{\rm Eq\,E1}]_{_{\rm G2D}}\rho\subF v\subF^{3/2}}{5v^{1/2}}\,.
\end{eqnarray}
Substituting this into Eq. (\ref{EQN:Deltat1}), the post-shock cooling time becomes
\begin{eqnarray}
\Delta t^{^{\rm PSC}}_{_{\rm DUST}}&\sim&\frac{5}{2\,[{\rm Eq\,E1}]_{_{\rm G2D}}\,\rho\subF\,v\subF}
\end{eqnarray}
Comparing this with Eq. (\ref{EQN:Deltat_COLOW}), and noting that $[{\rm Eq\,E1}]_{_{\rm G2D}}\!\ll\![{\rm Eq\,D4}]_{_{\rm CO.LOW}}$, we conclude that dust makes a negligible contribution to post-shock cooling, in the density regime with which we are concerned (although, at even higher densities, $n_{{\rm H}_2}> 10^8\,{\rm cm}^{-3}$, dust might make a significant contribution to post-shock cooling, because line cooling would be optically thick).

\subsection{The compressional heating rate in a collapsing prestellar core}\label{SEC:COREHEAT}

The freefall time is
\begin{eqnarray}\label{EQN:FREEFALL}
t_{_{\rm FF}}&=&\left(\frac{3\pi}{32G\rho}\right)^{1/2}\,.
\end{eqnarray}
It follows that the compressional heating rate in a collapsing prestellar core is of order
\begin{eqnarray}\nonumber
\Gamma_{_{\rm COMP}}\!&\!\sim\!&\!\frac{u_{_{\rm THERM}}}{t_{_{\rm FF}}}\;\sim\;\frac{3\rho a^2}{2}\!\left(\frac{32G\rho}{3\pi}\right)^{\!1/2}\\\label{EQN:CORECOOL}
\!&\!\sim\!&\!\left[7.14\!\times\! 10^{-4}\,{\rm g}^{-1/2}\,{\rm cm}^{3/2}\,{\rm s}^{-1}\right]_{_{\rm COMP}}\rho^{3/2}a^{2}\!.
\end{eqnarray}

\subsection{CO cooling of a collapsing prestellar core}\label{SEC:CORE_CO}

Since we are concerned here with condensations that are marginally Jeans unstable, we adapt the high-density, optically thick CO cooling expression (Eq. \ref{EQN:CO_HIGH1}) to the properties in the interior of a critical Bonnor-Ebert sphere. This means that, where we need an estimate for $\Delta v/L$, we must substitute $\Delta v\!\rightarrow\sigma_{_{\rm NT}}$ and $L\!\rightarrow\!R_{_{\rm BE}}\!\sim\!\xi_{_{\rm BE}}((a\subL^2\!+\!\sigma_{_{\rm NT}}^2)/4\pi G\rho\subC)^{1/2}$. Here, $\sigma_{_{\rm NT}}$ is the non-thermal contribution to the one-dimensional radial velocity dispersion; $R_{_{\rm BE}}$ and $\xi_{_{\rm BE}}\!=\!6.45$ are, respectively, the physical and dimensionless radii of a critical Bonnor-Ebert sphere; and $\rho_{_{\rm CENTRE}}$ is the central density \citep[see][]{Chandras1939, ChanWare1949}. The non-thermal velocity dispersions in low-mass marginally Jeans unstable prestellar cores are observed to be trans- or sub-sonic, so we put $\sigma_{_{\rm NT}}\!\sim\! a\subL$. The high-density optically thick CO line-cooling rate (Eq. \ref{EQN:CO_HIGH1}) then becomes
\begin{eqnarray}\label{EQN:CO_HIGH3}
\Lambda_{_{\rm CO.HIGH}}&\sim&[{\rm Eq}\,\ref{EQN:CO_HIGH4}]'_{_{\rm CO.HIGH}}\,\rho\subC^{1/2}\,a\subL^8\,,
\end{eqnarray}
with
\begin{eqnarray}\nonumber
[{\rm Eq}\,\ref{EQN:CO_HIGH4}]'_{_{\rm CO.HIGH}}&\sim&\frac{(2\pi G)^{1/2}[{\rm Eq\,D6}]_{_{\rm CO.HIGH}}}{6.45}\\\nonumber
&\sim&\left[1.7\times 10^{-48}{\rm g}^{1/2}{\rm cm}^{-15/2}{\rm s}^5\right]'_{_{\rm CO.HIGH}}\!.\\\label{EQN:CO_HIGH4}
\end{eqnarray}

In a prestellar core, low-density optically thin CO cooling (Eq. \ref{EQN:CO_LOW1}) dominates over high-density optically thick CO cooling (Eq. \ref{EQN:CO_HIGH3}), if
\begin{eqnarray}\label{EQN:THIN2THICK;CORE}
n_{_{{\rm H}_2}} &\la&1.1\times 10^3\,{\rm cm}^{-3}\,\left(\frac{T\subL}{10\,{\rm K}}\right)^{5/3}\,.
\end{eqnarray}
At these relatively low densities, optically thin CO cooling (Eq. \ref{EQN:CO_LOW1}) delivers the required cooling rate (Eq. \ref{EQN:CORECOOL}) provided
\begin{eqnarray}\label{EQN:THINCORE1}
n_{_{{\rm H}_2}} &\ga&1.5\,{\rm cm}^{-3}\,\left(\frac{T\subL}{10\,{\rm K}}\right)^{-1}\,,
\end{eqnarray}
which is easily satisfied. At higher densities, optically-thick CO cooling (Eq. \ref{EQN:CO_HIGH3}) delivers the required cooling rate (Eq. \ref{EQN:CORECOOL}), provided
\begin{eqnarray}\label{EQN:THICKCORE2}
n_{_{{\rm H}_2}} &\la&3.1\times 10^4\,\left(\frac{T\subL}{10\,{\rm K}}\right)^3\,,
\end{eqnarray}
and this is a critical constraint. We note that the density and temperature dependence in Eq. (\ref{EQN:THICKCORE2}) defines a minimum  Jeans mass, which is universal in the sense that it is independent of $n_{_{{\rm H}_2}}$ and $T\subL$. Specifically, and irrespective of the temperature, $T\subL$, $^{12}$C$^{16}$O cannot on its own provide sufficient cooling for cores with mass $M\la2.7\,{\rm M}_{_\odot}$ ($\ga 80\%$ of cores) to condense out. For cores with $M>2.7\,{\rm M}_{_\odot}$ it can only provide sufficient cooling for the core to condense to a fraction 
\begin{eqnarray}
f&\sim&\left(\!\frac{2.7\,{\rm M}_{_\odot}}{M}\!\right)^{2/3}
\end{eqnarray}
of its original size (for example, to $\sim 0.42$ of its original size for a $10\,{\rm M}_{_\odot}$ core).

In reality there are contributions from ${\cal N}_{_{\rm MOL}}$ other molecules (or atoms or ions), such as the isotopomers of CO, so these limits might be reduced somewhat. If we put ${\cal N}_{_{\rm MOL}}\!=5$, molecular lines cannot provide sufficient cooling for cores with mass $M\la 1.2\,{\rm M}_{_\odot}$ ($\ga 55\%$ of cores), and a $10\,{\rm M}_{_\odot}$ core can only contract to $\sim 0.24$ of its original size before line cooling ceases to cope.

However, CO (and other molecules) are starting to freeze out onto dust at the high densities in low- and intermediate-mass  prestellar cores (see Section \ref{SEC:BasTherm} and Appendix \ref{APP:TIMESCALES}). The combination of freeze-out and optical depth makes line cooling increasingly ineffective in a collapsing prestellar core, whilst the cooling requirements of collapsing prestellar cores are becoming ever more demanding.

\subsection{Dust cooling of a collapsing prestellar core}

Comparing Eqs. (\ref{EQN:GAS2DUST}) and (\ref{EQN:CORECOOL}), we find that dust cooling can deliver sufficient cooling for a prestellar core to collapse, provided
\begin{eqnarray}\label{EQN:THINCORE1}
n_{_{{\rm H}_2}} &\ga&2\times 10^4\,{\rm cm}^{-3}\,\left(\frac{T\subL}{10\,{\rm K}}\right)^{-1}\,.
\end{eqnarray}

Clearly dust cooling is much better able to cool a collapsing prestellar core than CO, but only if the density is high enough. Specifically we require that the ram-pressure of the gas flowing into the shock-compressed layer, $P_{_{\rm RAM}}=\rho\subF v\subF^2$, delivers a sufficiently high hydrostatic pressure in the layer, $\rho\subL a\subL^2$, to satisfy Eq. (\ref{EQN:THINCORE1}), or
\begin{eqnarray}\nonumber
\frac{P}{\bar{m}_{_{{\rm H}_2}}}&\equiv&n_{_{{\rm H}_2{\rm .FLOW}}} v\subF^2\\
&\ga&\frac{P_{_{\rm CRIT}}}{\bar{m}_{_{{\rm H}_2}}}\;\,\sim\;\,800\,{\rm cm}^{-3}({\rm km}/{\rm s})^2\,,
\end{eqnarray}
independent of the temperature.

The surface-density of the layer when it fragments is then
\begin{eqnarray}
\Sigma\;\ga\;\Sigma_{_{\rm CRIT}}\!\!&\!\!\sim\!\!&\!\!\left(\!\frac{2 P_{_{\rm CRIT}}}{\pi G}\!\right)^{\!1/2}\,\sim\,100\,{\rm M}_{_\odot}\,{\rm pc}^{-2}\,,
\end{eqnarray}
which corresponds to a column-density
\begin{eqnarray}
N_{_{{\rm H}_2}}\;\ga\;N_{_{{\rm H}_2{\rm .CRIT}}}\!&\!=\!&\!\frac{\Sigma_{_{\rm CRIT}}}{\bar{m}_{_{{\rm H}_2}}}\;\sim\;4\times 10^{21}\,{\rm H}\,{\rm cm}^{-2}\,,
\end{eqnarray}
and visual extinction\footnote{We convert column-density into visual extinction using the analytic evaluation of $\kappa\subD$ (Eqn. \ref{EQN:kappaD2}) and $Q_{_{\rm V}}\!=\!1.5$, for internal consistency, since $\kappa\subD$ also enters into the expression for dust cooling (Eqn. \ref{EQN:G2D1}). If we were to reduce $\kappa_{_{\rm DUST}}$, so as to reproduce the \citet{Bohletal1978} conversion factor (i.e. $A_{_{\rm V}}=2N_{_{{\rm H}_2}}/(5.8\times 10^{21}\,{\rm cm}^{-2}\,R)\sim 5\,{\rm mag}$, where $R\!=\!A_{_{\rm V}}/E_{_{\rm B-V}}\!\sim\!3.2$ is the ratio of total visual to selective extinction), we would also need to increase the critical column-density, in such a way that the critical extinction would be unchanged.}
\begin{eqnarray}
A_{_{\rm V}}\;\,\ga\;\,A_{_{\rm V.CRIT}}\!&\!=\!&\!\frac{5\Sigma_{_{\rm CRIT}}\kappa\subD Q_{_{\rm V}}}{\ln(100)}\;\,\sim\;\,9\,{\rm mag.}
\end{eqnarray}

The diameter of a filament when it {\it starts} to fragment into prestellar cores is given by Eq. (\ref{EQ:R_FIL_1}),
\begin{eqnarray}
2R_{_{\rm FIL}}&\la&0.16\,{\rm pc}\,\left(\frac{T\subL}{10\,{\rm K}}\right)\,,
\end{eqnarray}
and the line density of a filament is the product of $\dot{\mu}_{_{\rm IN}}$ (Eq. \ref{EQN:mudot}) and $3t_{_{\rm CRIT}}$ (Eq. \ref{EQN:tCRIT1}), i.e.
\begin{eqnarray}
\mu_{_{\rm FIL}}&\sim&\frac{3\,a\subL^2}{2\,G}\;\,\simeq\;\,13\,{\rm M}_{_\odot}\,{\rm pc}^{-2}\,\left(\frac{T\subL}{10\,{\rm K}}\right)\,.
\end{eqnarray}
\citet{HennAndr2013} propose an alternative explanation for the characteristic width of filaments, involving the balance between turbulence-driven accretion and dissipation of turbulence by ion-neutral friction.

At their inception, the cores typically have radii and masses corresponding to a critical Bonnor-Ebert sphere, with mean density equal to $\rho_{_{\rm CRIT}}$, viz.
\begin{eqnarray}\nonumber
2R_{_{\rm CORE}}&\sim&\left(\frac{3\,\mu_{_{\rm B}}}{\pi\, G\,\rho_{_{\rm CRIT}}\,\xi_{_{\rm B}}}\right)^{1/2}\,a\subL\\
&\sim&0.12\,{\rm pc}\,\left(\frac{T\subL}{10\,{\rm K}}\right)\,,\\\nonumber
M_{_{\rm CORE}}&\sim&\left(\frac{3\,\mu_{_{\rm B}}^3}{4\,\pi\,G^3\,\rho_{_{\rm CRIT}}\,\xi_{_{\rm B}}^3}\right)^{1/2}\,a\subL^3\\
&\sim&1\,{\rm M}_{_\odot}\,\left(\frac{T\subL}{10\,{\rm K}}\right)^2\,.
\end{eqnarray}
Here $\mu_{_{\rm B}}=15.7$ and $\psi_{_{\rm B}}=2.6$ are obtained from the Isothermal Function \citep[see Appendix \ref{APP:BONNOR} and][]{Chandras1939,ChanWare1949}. 

\section{Discussion and conclusions}\label{SEC:CONC}

There are three critical aspects of the model we have developed here: (i) the role of the early fragmentation mode, which breaks the layer into filaments, whilst the layer is still accumulating; (ii) fact that the gas then rearranges itself -- driven by self-gravity, but at approximately constant density -- into filaments and then cores, before going into collapse; and (iii) the role of cooling and the consequences of the switch from line cooling (which regulates the post-shock cooling near the surface of the layer) to dust cooling (which facilitates core collapse once the material in a layer has been assembled into cores). We briefly discuss each of these in turn. In addition, we note how the results are changed if the metallicity, and hence the dust abundance, are different. Finally we summarise our main conclusions.

\subsection{The early and late fragmentation modes for a shock-compressed layer}\label{SEC:ReduDime}

Most previous work on fragmentation in configurations of reduced dimension (i.e. not standard Jeans fragmentation of an extended three-dimensional medium) has tended to start from the premise that a layer or filament is essentially through with assembling, and is also close to a plane-parallel or cylindrical equilibrium configuration, before it fragments gravitationally. This scenario has the advantage that an equilibrium state can be defined, to which one can then apply perturbation analysis. However, it appears to involve an internal  contradiction, namely that the timescale for gravitational fragmentation of the assembled layer or filament is much shorter than the timescale on which the layer or filament is assembled. Therefore the premise that the layer or filament waits patiently to be assembled before trying to fragment gravitationally is untenable. Rather, the layer or filament will attempt to fragment while it is being assembled. In contrast, a key assumption of the model proposed here is that gravitational fragmentation of a shock-compressed layer occurs while matter continues to flow into the layer; the layer tries to fragment at the same time as it accumulates, and non-linear fragmentation gets going once the timescale for fragmentation becomes shorter than the time for which the layer has been accumulating. It is appropriate that we should justify this assumption.

To do so, we define characteristic length- and time-scales, $L\subO$ and $t\subO$, in terms of the pre-shock density, $\rho\subF$, and the post-shock sound speed, $a\subL$,
\begin{eqnarray}
L\subO&=&\frac{a\subL}{(G\rho\subF)^{1/2}}\,,\\
t\subO&=&\frac{1}{(G\rho\subF)^{1/2}}\,;
\end{eqnarray}
and we recall that the inflow velocity, $v\subF$, can be expressed in terms of the sound speed, $a\subL$, using the Mach Number, 
\begin{eqnarray}
{\cal M}&=&\frac{v\subF}{a\subL}\,. 
\end{eqnarray}

In our model for {\it early} gravitational fragmentation of a layer into filaments, while the layer is still accumulating, the timescale on which non-linear fragmentation develops, $t_{_{\rm FRAG.EARLY}}$, is the same as the time it takes for the layer to accumulate sufficient column-density for the fragmentation instability to become non-linear, $t_{_{\rm ACCUM.EARLY}}$. The surface density of the layer increases with time, and consequently the timescale on which the fastest growing early gravitational mode develops decreases with time. Non-linear fragmentation occurs when the decreasing fragmentation timescale becomes comparable with the time for which the layer has been accumulating. This is the time at which the competition between early gravitational fragmentation of the layer, and growth of the surface-density of the layer, tips in favour of fragmentation. From Eq. (\ref{EQN:tCRIT1}), and ignoring purely numerical factors (since they are not robust), we have
\begin{eqnarray}
t_{_{\rm FRAG.EARLY}}&\sim &t\subO{\cal M}^{-1/2}\,, \\
t_{_{\rm ACCUM.EARLY}}&\sim &t\subO{\cal M}^{-1/2}\,.
\end{eqnarray}
The typical separation between filaments (Eq. \ref{EQN:2YCRIT}) is
\begin{eqnarray}
2Y_{_{\rm FRAG.EARLY}}&\sim&L\subO{\cal M}^{-1/2}\,,
\end{eqnarray}
but the thickness of the layer (Eq. \ref{EQN:Thickness}) is much smaller,
\begin{eqnarray}
2Z_{_{\rm FRAG.EARLY}}&\sim&L\subO{\cal M}^{-3/2}\,.
\end{eqnarray}

In contrast, the {\it late} mode of layer fragmentation analysed by \citet{LarsonRB1985} can only occur when the layer is sufficiently massive and thick (approximately, thicker than a Jeans wavelength, {\it as evaluated at the density in the layer}, $\rho\subF{\cal M}^2$). This means that the thickness of the layer must be 
\begin{eqnarray}
2Z_{_{\rm FRAG.LATE}}&\sim&L\subO{\cal M}^{-1}\,,
\end{eqnarray}
and it takes a time 
\begin{eqnarray}
t_{_{\rm ACCUM.LATE}}&\sim &t\subO\,,
\end{eqnarray}
for the layer to grow this big. If this stage were ever reached, the time required for the late fragmentation mode to develop would -- by comparison -- be very short,
\begin{eqnarray}
t_{_{\rm FRAG.LATE}}&\sim &t\subO{\cal M}^{-1}\,.
\end{eqnarray}
However, the layer is very unlikely to reach this stage intact, because -- since $\;t_{_{\rm ACCUM.EARLY}}<t_{_{\rm ACCUM.LATE}}$ -- it is likely to have already undergone early fragmentation into filaments.

\begin{figure}
\begin{center}
\includegraphics[width=1.80\columnwidth,angle=-90]{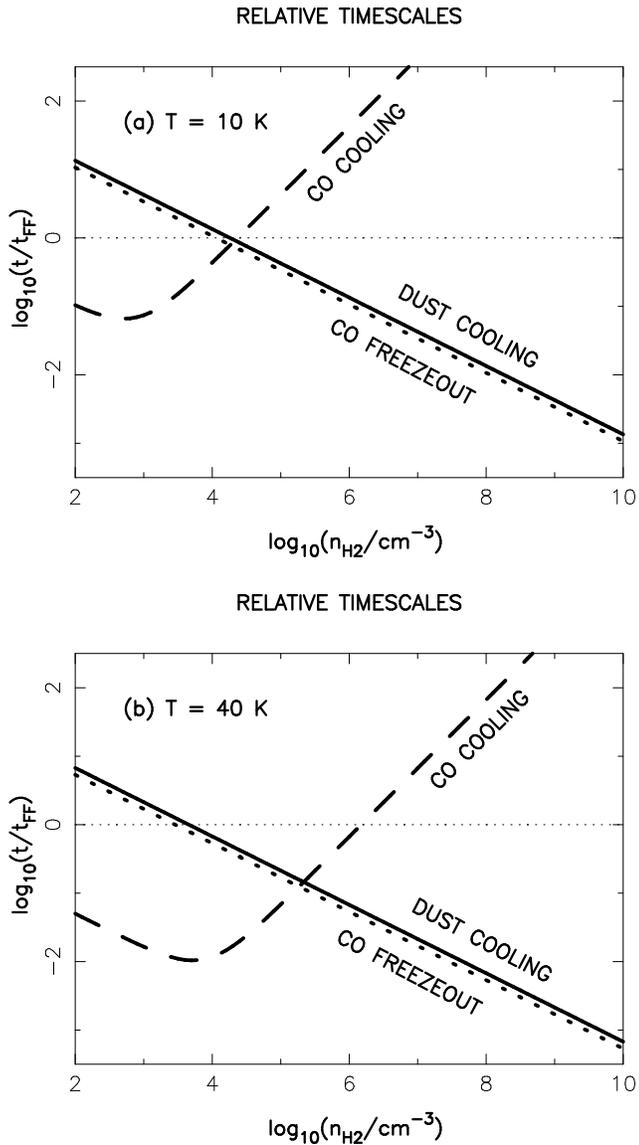}
\caption{Timescales for cooling and molecular freeze-out, relative to freefall, for marginally Jeans unstable cores, as a function of density. The abscissa is the logarithm of the number-density of molecular hydrogen, $\log_{_{10}}(n_{_{{\rm H}_2}}/{\rm cm}^{-3})$. The ordinate is the logarithm of the timescale, in units of the freefall time, for dust cooling (thick full line), CO cooling (thick dashed line), and CO freeze-out (thick dotted line). The thin dotted horizontal line indicates where the timescales are equal to the freefall time. CO cooling can only support core collapse and fragmentation at densities where the CO-cooling line is below the thin horizontal line and the CO-freeze-out line is above it. Dust cooling can only support core collapse and fragmentation at densities where the dust-cooling line is below the thin horizontal line. (a) $T\!=\!10\,{\rm K}$. (b) $T\!=\!40\,{\rm K}$.}
\label{FIG:Cooling}
\end{center}
\end{figure}

\subsection{The constant-density phase}

Between (a) shock compression (when gas flows into the accretion shock bounding the layer, and then undergoes post-shock cooling facilitated by line cooling) and (c) core collapse (when the gas heads for stardom, facilitated by dust cooling), there is an interim (b) during which the gas is redistributed, under the influence of self-gravity, but at approximately constant density. This is the phase during which the layer fragments into filaments, and the filaments fragment into cores. The density is approximately constant because the pressure in the layer and embedded filaments is approximately equal to the ram pressure of the gas that continues to flow into the layer and embedded filaments. Because the density is approximately constant, there is little compressional heating, and therefore the gas can stay cool; the radiative cooling rate (due to lines and/or dust) can easily match any heating due to external agents like UV photons and cosmic rays. Only when the gas has been collected into cores does it start to collapse to much higher (ultimately stellar) densities. Thus the shock plus post-shock cooling due to line emission, which delivers the gas to sufficient density to couple thermally to the dust, occurs some time, $\sim 3t_{_{\rm FRAG.EARLY}}\sim (a\subL/G\rho\subF v\subF)^{1/2}$, before the gravitational collapse and fragmentation that requires this density to enable dust cooling. This extended period at constant density is also a time during which CO molecules are likely to freeze out onto dust grains, so that they are not available to help the gas cool if it subsequently becomes part of a collapsing core. We do not include this consideration in our analysis, but it can only make stronger our conclusion that dust cooling is critical during the core collapse.

\subsection{The basic radiative cooling requirements for turbulent fragmentation}\label{SEC:BasTherm}

From a radiative cooling viewpoint, there are two critical junctures in the model for turbulent fragmentation that we have developed here.

The first critical juncture is when gas streams collide to form a shock-compressed layer. Significant compression, i.e. a significant increase in density, requires that the post-shock gas can radiate efficiently and cool down quickly to a low temperature, so that most of the gas in the layer is cold and dense, and therefore susceptible to gravitational instability. The results presented in Section \ref{SEC:SHOCK_CO} indicate that, for the conditions obtaining in local molecular clouds, CO is a very effective post-shock coolant. In the hot rarefied gas immediately behind the shock, high-$J$ lines of CO are excited, and there is a wide range of velocities. Both of these effects increase the integrated line emissivity. In contrast, the density is in general too low for efficient thermal coupling between the gas and the dust, and consequently dust contributes little to post-shock cooling (see Section \ref{SEC:PS.DUST}).

The second critical juncture is after some of the gas in the layer has been assembled into prestellar cores -- by lateral flows in the layer, into and then along, filaments, at approximately constant density -- and this gas starts to condense out due to its self-gravity. By this stage, many molecular cooling lines have become optically thick, so their ability to radiate away the gravitational potential energy being released is limited. In addition, many of the line coolants have also frozen out onto the dust, and are therefore no longer able to emit line radiation. However, by this stage the density is so high that the gas is strongly thermally coupled to the dust, and the dust is able to emit across a very broad range of continuum wavelengths. Furthermore, the dust emission at these wavelengths is not yet optically thick --- except for cores near the opacity limit, i.e. cores with a few Jupiter masses --- so the radiation escapes easily. Additionally any molecules that have frozen out onto the dust serve simply to increase the surface area of the dust, thereby enhancing further its ability to radiate at long wavelengths.

Fig. \ref{FIG:Cooling} shows how the timescales for cooling and CO-freeze-out depend on density, for a marginally Jeans unstable core. The timescales are given as a function of the freefall time -- using the expressions derived in Appendix \ref{APP:TIMESCALES} -- so they are equal on the thin horizontal line. The thick dashed line is for CO cooling, the thick dotted line is for CO freeze-out, and the thick full line is for dust cooling. CO cooling can only support core collapse and fragmentation at densities where the CO freeze-out line is above the thin horizontal line (so CO actually exists in the gas phase), and where the CO cooling line is below the thin horizontal line (so the CO cooling can cope with the thermal energy being generated by compression). Dust cooling can only support core collapse and fragmentation at densities where the dust cooling line is below the thin horizontal line (so dust cooling can cope with the thermal energy being generated by compression).

At $T\!=\!10\,{\rm K}$ (upper plot on Fig. \ref{FIG:Cooling}), and densities above $n_{_{{\rm H}_2}}\!\sim\!2\times 10^4\,{\rm cm}^{-3}$, the CO cooling line is above zero, i.e. the CO cooling timescale is longer than the freefall timescale and CO cooling is unable to support core collapse and fragmentation. In addition, the CO freeze-out line is below zero, i.e. the CO freeze-out timescale is shorter than the freefall timescale and the CO is rapidly disappearing anyway; in fact CO has probably been freezing out all the time the gas has been collecting into filaments, and then into cores. In contrast, the dust cooling line is below zero for densities greater than $n_{_{{\rm H}_2}}\!\sim\!2\times 10^4\,{\rm cm}^{-3}$, hence the dust cooling timescale is shorter than the freefall timescale and dust cooling is well able to support core collapse and fragmentation at these densities.

At $T\!=\!40\,{\rm K}$ (lower plot on Fig. \ref{FIG:Cooling}), CO cooling is more efficient, and the CO cooling timescale remains below the freefall timescale up to densities $n_{_{{\rm H}_2}}\!\sim\!10^6\,{\rm cm}^{-3}$. However, CO freeze-out also occurs more quickly at these higher temperatures, and the timescale for CO freeze-out drops below the freefall timescale at densities greater than $n_{_{{\rm H}_2}}\!\sim\!10^4\,{\rm cm}^{-3}$, so CO cooling cannot support core collapse and fragmentation beyond this density. Dust cooling is also more efficient, and the timescale for dust cooling falls below the freefall timescale for densities greater then $n_{_{{\rm H}_2}}\!\sim\!5\times 10^3\,{\rm cm}^{-3}$, so dust cooling is well able to support core collapse and fragmentation at these densities.

It is the switch from molecular line cooling (which is becoming increasingly ineffective) to dust continuum cooling (which is becoming increasingly effective) that defines a critical ram-pressure for the turbulent flows producing compression -- and hence a critical surface-density, a critical diameter for filaments, and critical diameters and masses for prestellar cores -- even though the gas only avails itself of this efficient dust cooling some time after it has been compressed to the required density, because there is an interim, between shock compression at the boundary of the layer and core collapse, during which the gas assembles itself into embryonic cores at approximately constant density.

\subsection{The dependence on metallicity}

At the level of approximation in our analysis, it would be inappropriate to consider variations in the nature of dust with metallicity. Therefore we simply assume that the coefficient for dust cooling, $[{\rm Eq\,E1}]_{_{\rm G2D}}$ is proportional to the abundance of dust by mass, $Z\subD$. The critical ram-pressure (Eq. \ref{EQN:PCrit1}) varies as $Z\subD^{-2}$; the critical surface-density, column-density and extinction (Eqs. \ref{EQN:SigmaCrit}, \ref{EQN:NH2Crit} and \ref{EQN:AVCrit}) all vary as $Z\subD^{-1}$; and the characteristic diameters of cores and filaments, and the characteristic mass of cores (Eqs. \ref{EQN:2RCrit} and \ref{EQN:MCrit}) all vary as $Z\subD$. In the conclusions that follow, we explicitly spell out these dependences.

\subsection{Conclusions}

We have developed a simple analytic model for the basic physical processes involved in turbulent fragmentation, concentrating on the thermodynamic and geometric aspects, and on the conditions prevailing in local star formation regions. Our main conclusions are as follows.

\begin{itemize}

\item In the radiative cooling regions behind the shocks where turbulent flows collide, line cooling by molecules (or atoms or ions) dominates over dust cooling, because the density is too low for the gas to transfer much thermal energy to the dust, whereas many lines are excited, and they are broadened by the range of velocities spanned by the decelerating gas. Line cooling quickly reduces the gas temperature, so that the bulk of the shocked gas in the resulting dense layer is cold and approximately isothermal. 

\item After the gas passes through this cooling region, it initially has approximately uniform density, because firstly the shocked region is contained by the ram-pressure of the inflowing gas, secondly self-gravity plays only a minor role in holding the layer together at these early times (the surface-density is too low), and thirdly there is sufficient time (many sound-crossing times) for the gas to approach its hydrostatic equilibrium configuration, and for turbulent motions to decay. Therefore the pressure in the shocked layer is approximately uniform, and approximately equal to the ram-pressure of the inflowing gas.

\item Paradoxically, the layer fragments gravitationally into filaments before self-gravity ever becomes strong enough to make a significant contribution to holding the layer together. This is because self-gravity {\it is} important on relatively long wavelengths parallel to the plane of the layer, breaking it up into filaments --- even though it {\it is not} important perpendicular to the plane of the layer, and contributes very little to holding the layer together. A consequence of this is that the layer is unlikely to become massive enough, {\it either} to require an equilibrium in which there is a steep density gradient perpendicular to the mid-plane, providing support against self-gravity, {\it or} to fragment via the late mode analysed by \citet{LarsonRB1985}.

\item If there is a large-scale magnetic field with a significant component in the plane of the layer, or if the colliding flows deliver significant bulk angular momentum to the layer, the filaments will tend to be aligned (perpendicular to the field and/or the angular momentum), and cores will condense out of the filament with roughly even spacing. If there is not, then filaments will be randomly oriented, forming a network, with the larger cores tending to condense out at the intersections of filaments.

\item The spacing between filaments gives a lower limit on the Mach Number, ${\cal M}=v\subF/a\subL$, of the shock producing the layer from which the filaments formed.

\item Once the gas has accreted onto the layer and cooled down, it flows laterally into filaments, and then along the filaments into prestellar cores, at approximately constant density; as explained above, the density is controlled by the ram pressure of the inflow that continues to feed the layer.

\item This phase, during which some of the shocked gas in the layer assembles itself into cores, takes a time $\sim (a\subL/G\rho\subF v\subF)^{1/2}$, and the cooling molecules are likely to start freezing out during this phase.

\item Even if there has been little freeze-out, molecular-line cooling struggles to radiate away the gravitational potential energy released by a condensing low-mass prestellar core, whereas dust cooling can cope easily in this situation. This is because the density in a low-mass prestellar core is high enough for the gas and dust to be thermally coupled, and dust cooling delivers emission across the full bandwidth of the continuum.

\item This thermal coupling requirement defines a critical (and temperature independent) ram-pressure for the gas flowing into the layer, $P_{_{\rm CRIT}}\sim 4\times 10^{-11}\,{\rm dynes}$, and a critical surface-density, $\Sigma_{_{\rm CRIT}}\sim 100\,{\rm M}_{_\odot}\,{\rm pc}^{-2}$, that must be delivered by the colliding flows if the layer is to fragment efficiently. Flows that have insufficient ram-pressure (because they are too rarefied or too slow), or deliver insufficient column-density (because they peter out), produce layers which may break up into filaments, but are less likely to spawn stars. In the notation of \citet{Hacaetal2013}, they tend to be sterile.

\item The critical surface-density corresponds to a critical column-density, $N_{_{{\rm H}_2.{\rm CRIT}}}\sim 4\times 10^{21}\,{\rm H}_2\,{\rm cm}^{-2}$ and a critical visual extinction, $A_{_{\rm V.CRIT}}\sim 9\,{\rm mag}$.

\item It also corresponds to a characteristic diameter and line-density for filaments, and a characteristic diameter and mass for prestellar cores, \\\\
\indent$2R_{_{\rm FIL}}\sim 0.12\,{\rm pc}\,(T\subL/10{\rm K})$, \\
\indent$\mu_{_{\rm FIL}}\sim 13\,{\rm M}_{_\odot}\,{\rm pc}^{-1}\,(T\subL/10{\rm K})$, \\
\indent$2R_{_{\rm CORE}}\sim 0.12\,{\rm pc}\,(T\subL/10{\rm K})$, \\
\indent$M_{_{\rm CORE}}\sim 1.3\,{\rm M}_{_\odot}\,(T\subL/10{\rm K})^2\;$. \\\\
Here $T\subL$ is the {\it effective} temperature in the layer (i.e. a temperature that may need to be augmented to take into account any non-thermal pressure, due to residual turbulence, magnetic fields, etc.). 

\item These values of $\Sigma_{_{\rm CRIT}}$, $N_{_{{\rm H}_2.{\rm CRIT}}}$, $A_{_{\rm V.CRIT}}$, $2R_{_{\rm FIL}}$, $\mu_{_{\rm FIL}}$, $2R_{_{\rm CORE}}$ and $M_{_{\rm CORE}}$, are very close to those observed in local star formation regions. 

\item If this model is also applicable to star formation in regions with a different fractional abundance by mass of dust, $Z\subD$, the critical and characteristic quantities become \\\\
\indent$P_{_{\rm CRIT}}\sim 4\times 10^{-11}\,{\rm dynes}(Z\subD/0.01)^{-2}$, \\
\indent$\Sigma_{_{\rm CRIT}}\sim 100\,{\rm M}_{_\odot}\,{\rm pc}^{-2}(Z\subD/0.01)^{-1}$, \\
\indent$N_{_{{\rm H}_2.{\rm CRIT}}}\sim 4\times 10^{21}\,{\rm H}_2\,{\rm cm}^{-2}(Z\subD/0.01)^{-1}$, \\
\indent$A_{_{\rm V.CRIT}}\sim 9\,{\rm mag}(Z\subD/0.01)^{-1}$, \\
\indent$2R_{_{\rm FIL}}\sim 0.12\,{\rm pc}\,(T\subL/10{\rm K})\,(Z\subD/0.01),$ \\
\indent$\mu_{_{\rm FIL}}\sim 13\,{\rm M}_{_\odot}\,{\rm pc}^{-1}\,(T\subL/10{\rm K})$, \\
\indent$2R_{_{\rm CORE}}\sim 0.12\,{\rm pc}\,(T\subL/10{\rm K})\,(Z\subD/0.01)$, \\
\indent$M_{_{\rm CORE}}\sim 1.3\,{\rm M}_{_\odot}\,(T\subL/10{\rm K})^2\,(Z\subD/0.01)$. \\\\
In locations where the dust abundance is lower, the critical values of pressure, surface-density, column-density and extinction are larger.\footnote{There is some evidence for this trend in Fig. 2 of \citet{Krumholz2014}, which shows that in a lower-metallicity system like the Small Magellanic Cloud, efficient star formation requires a larger surface-density of interstellar gas. However, we should note that this plot is concerned with surface densities averaged over quite large regions of external galaxies, and so the threshold where large-scale star formation becomes much more efficient on Krumholz's plot is lower than the local value we calculate -- for example, around $\sim 5\,{\rm M}_{_\odot}\,{\rm pc}^{-2}$ in Milky Way type galaxies.} The diameter of a filament, and the diameter and mass of a critical core are all smaller, unless this is compensated by the gas being hotter, due to a higher background radiation field, and/or a lower abundance of coolants, i.e. lower metallicity.

\end{itemize}

\section*{Acknowledgements}

APW gratefully acknowledges the support of a consolidated grant (ST/K00926/1) from the UK Science and Technology Funding Council, and thanks Seamus Clarke, Oliver Lomax, Paul Clark and Nicolas Peretto for useful discussions.

\appendix

\section{The gravitational acceleration amplifying a corrugation wave in a uniform-density layer}\label{APP:gY}

To compute the gravitational acceleration amplifying a corrugation wave in a plane-parallel layer with boundaries at  $z=\pm Z\subL$, we assume that the corrugation has infinite extent in the $x$-direction, and that the wave-vector is in the $y$-direction. Thus, without loss of generality, the matter involved in a single corrugation, hereafter a proto-filament, is initially contained in a plank-shaped region, $-\infty\!<\!x\!<\!+\infty,\;-Y\!<\!y\!<\!+Y,\;-Z\subL\!<\!z\!<\!+Z\subL$, with uniform density, $\rho\subL$. It is plank-shaped in the sense that $Y\gg Z\subL$, as illustrated in Fig. 1a and demonstrated in Section \ref{SEC:DOMWL}.

The contribution to the $y$-component of the gravitational acceleration at $(x,y,z)\!=\!(0,-Y,0)$, from the two-dimensionally infinitesimal part of the proto-filament in $(-\infty\!<\!x\!<\!\infty;\,y,y+dy;\,z,z+dz)$, is
\begin{eqnarray}
d^2g_{_y}(-Y)&=&\frac{2\,G\,\rho\subL\,(Y+y)\,dy\,dz}{\left((Y+y)^2+z^2\right)}\,.
\end{eqnarray}
Integrating from $y\!=\!-Y$ to $y\!=\!+Y$, this gives
\begin{eqnarray}
dg_{_y}(-Y)&=&G\,\rho\subL\,dz\,\ln\!\left(\frac{4Y^2+z^2}{z^2}\right)\,.
\end{eqnarray}
Integrating from $z\!=\!-Z\subL$ to $z\!=\!+Z\subL$, we obtain
\begin{eqnarray}\nonumber
g_{_y}(-Y)\!&\!=\!&\!2G\rho\subL\left\{Z\subL\ln\!\left(\!\frac{4Y^2+Z\subL^2}{Z\subL^2}\!\right)\right. \\
&&\hspace{2.25cm}\left.+4Y\tan^{-1}\!\left(\!\frac{Z\subL}{2Y}\!\right)\right\}\!.
\end{eqnarray}

To zeroth order in $Z\subL/Y$, this reduces to
\begin{eqnarray}\label{EQN:gy1}
g_{_y}(-Y)\,\simeq\,4G\rho\subL Z\subL\left\{1+\ln\!\left(\frac{2Y}{Z\subL}\right)\right\}.
\end{eqnarray}
In Section \ref{SEC:DOMWL} we show that $Y/Z\subL\simeq v\subF/2^{1/2}a\subL$ (Eq. \ref{EQN:Y2Z}), so for $a\subL=0.2\,{\rm km}\,{\rm s}^{-1}$ (our fiducial isothermal sound speed) and $1\,{\rm km}\,{\rm s}^{-1}\la v\subF\la7\,{\rm km}\,{\rm s}^{-1}$ (typical bulk velocities in a turbulent molecular cloud), we have $2\la\ln(2Y/Z\subL)\la 4$. Therefore we set the braces on the righthand side of Eq. (\ref{EQN:gy1}) to $\{4\!\pm\! 1\}$; the resulting error on $g_{_y}$ is always less than $25\%$, and in all the expressions that derive from $g_{_y}$ the error is less than $12\%$. Substituting $2\rho\subL Z\subL\simeq\Sigma\subL$, we obtain
\begin{eqnarray}\label{EQN:gy2}
g_{_y}(-Y)&\simeq&8\,G\,\Sigma\subL\,.
\end{eqnarray}
This is the result we use to compute the growth rates of corrugation waves in shock-compressed layers in Section \ref{SEC:CONDPF}.

\section{Isothermal filaments in hydrostatic balance}\label{APP:CYL}

The density profile of an infinite, cylindrically symmetric, self-gravitating, isothermal filament in hydrostatic balance is
\begin{eqnarray}\label{EQN:rhoFIL}
\rho(r)&=&\rho(0)\,\left(1\,+\,\frac{\pi \rho(0)r^2}{\mu_{_{\rm MAX}}}\right)^{\!-2}\,,
\end{eqnarray}
where 
\begin{eqnarray}\label{EQN:muMAX}
\mu_{_{\rm MAX}}&=&\frac{2a\subL^2}{G}
\end{eqnarray}
is the line-density above which the filament collapses to a line \citep{Ostriker1964}.

If the filament is contained by an external pressure, $P_{_{\rm EXT}}$, it is truncated at finite radius $R$, but the density profile is still given by Eq. (\ref{EQN:rhoFIL}). The mass per unit length out to radius $R$ is
\begin{eqnarray}\nonumber
\mu(R)&=&\int\limits_{r=0}^{r=R}\,\rho(r)\,2\pi r\,dr\\\label{EQN:muR}
&=&\mu_{_{\rm MAX}}\left\{1\,-\,\left(1\,+\,\frac{\pi \rho(0)R^2}{\mu_{_{\rm MAX}}}\right)^{\!-1}\right\}\,;
\end{eqnarray}
the density at the boundary is
\begin{eqnarray}\label{EQN:rhoR}
\rho(R)&=&\rho(0)\left(1\,+\,\frac{\pi \rho(0)R^2}{\mu_{_{\rm MAX}}}\right)^{\!-2}\;\,=\;\,\frac{P_{_{\rm EXT}}}{a\subL^2}\,;
\end{eqnarray}
and hence the radius is
\begin{eqnarray}\label{EQN:FILRAD1}
R(\mu)&=&\left(\frac{G\mu (\mu_{_{\rm MAX}}-\mu)}{2\pi P_{_{\rm EXT}}}\right)^{1/2}\,.
\end{eqnarray}

\section{The Jeans mass}\label{APP:BONNOR}

We assume that a marginally Jeans unstable prestellar core can be modeled as a critical Bonnor-Ebert sphere. In terms of the Isothermal Function, $\psi(\xi)$ \citep[see][]{Chandras1939}, $\mu=\xi^2{\rm e}^{-\psi(\xi)}$ is the dimensionless mass interior to arbitrary dimensionless radius $\xi$. $\;\mu_{_{\rm B}}\!=\!15.7$ is the value of $\mu$ at the boundary, $\xi_{_{\rm B}}\!=\!6.45$, of the critical Bonnor-Ebert sphere, which is the equilibrium state for an isothermal gas cloud contained by the maximum external pressure for which there is an equilibrium state \citep{ChanWare1949}.

The physical radius and mass of the critical Bonnor-Ebert sphere are given in terms of the isothermal sound speed, $a\subO$, and the {\it central} density, $\rho_{_{\rm C}}$, by
\begin{eqnarray}\label{EQN:RBEC1}
R_{_{\rm BE}}&=&\frac{a\subO\xi_{_{\rm B}}}{(4\pi G\rho_{_{\rm C}})^{1/2}}\;\,=\;\,1.82\,G^{-1/2}\,\rho_{_{\rm C}}^{-1/2}\,a\subO\,,\\\label{EQN:MBEC1}
M_{_{\rm BE}}&=&\frac{a\subO^3\mu_{_{\rm B}}}{G(4\pi G\rho_{_{\rm C}})^{1/2}}\;\,=\;\,4.43\,G^{-3/2}\,\rho_{_{\rm C}}^{-1/2}\,a\subO^3\,.
\end{eqnarray}

If we eliminate the central density in terms of the {\it mean} density,
\begin{eqnarray}
\bar{\rho}&=&\frac{3\mu_{_{\rm B}}\rho_{_{\rm C}}}{\xi_{_{\rm B}}^3}\;\,=\;\,0.175\,\rho_{_{\rm C}}\,,
\end{eqnarray}
we obtain
\begin{eqnarray}\label{EQN:RBEM}
R_{_{\rm BE}}&=&0.761\,G^{-1/2}\,{\bar\rho}^{\;-1/2}\,a\subO\,,\\\label{EQN:MBEM}
M_{_{\rm BE}}&=&1.86\,G^{-3/2}\,{\bar\rho}^{\;-1/2}\,a\subO^3\,.
\end{eqnarray}
We use these equations to compute the diameters and masses of prestellar cores condensing out of a filament.

If instead we eliminate the central density in terms of the {\it boundary} density,
\begin{eqnarray}
\rho_{_{\rm B}}&=&\rho_{_{\rm C}}{\rm e}^{-\psi_{_{\rm B}}}\;\,=\;\,0.0712\,\rho_{_{\rm C}}\,,
\end{eqnarray}
we obtain
\begin{eqnarray}\label{EQN:RBEB}
R_{_{\rm BE}}&=&0.485\,G^{-1/2}\,\rho_{_{\rm B}}^{-1/2}\,a\subO\,,\\\label{EQN:MBEB}
M_{_{\rm BE}}&=&1.18\,G^{-3/2}\,\rho_{_{\rm B}}^{-1/2}\,a\subO^3\,.
\end{eqnarray}

\section{Line-cooling by CO} \label{APP:CO}

Consider a linear molecule with moment of inertia $I_{_{\rm MOL}}$, and rotational energy levels, 
\begin{eqnarray}
E_{_J}&=&\frac{J(J+1)\hbar^2}{2I_{_{\rm MOL}}}\,.
\end{eqnarray}
The levels are significantly excited up to 
\begin{eqnarray}\label{EQN:JMAX}
J_{_{\rm MAX}}&\sim&\frac{(2I_{_{\rm MOL}}f_{_{\rm EX}}(n_{_{{\rm H}_2}},T)k_{_{\rm B}}T)^{1/2}}{\hbar}\,.
\end{eqnarray}
Here $f_{_{\rm EX}}$ is a weak function of $n_{_{{\rm H}_2}}$ and $T$, so we can set it to a constant over a limited range of $n_{_{{\rm H}_2}}$ and $T$.

If the density is low, the level populations are not thermalised, most of the molecules sit in the ground state, and most collisional excitations are followed by radiative de-excitation. If the lines are also optically thin, the resulting photon escapes, and the thermal energy that caused the collisional excitation is lost. The number of lines that can be excited is roughly proportional to $T^{1/2}$ (see Eq. \ref{EQN:JMAX}), as are the rates of excitation and the mean energies of the emitted photons. Hence, over a limited range of density and temperature, the contribution to the cooling rate per unit volume from a given molecular species can be approximated by
\begin{eqnarray}
\Lambda_{_{\rm MOL.LOW}}&\propto&n_{_{\rm MOL}}n_{_{{\rm H}_2}}T^{3/2}\,.
\end{eqnarray}
For example, at low densities and temperatures, the results of \citet{GoldLang1978} for $^{12}$C$^{16}$O can be fit with
\begin{eqnarray}\label{EQN:RAREFIED;THIN}
\Lambda_{_{\rm CO.LOW}}\!\!&\!\!\simeq\!\!&\!\!\left[1.3\times 10^4\,{\rm cm}^2\,{\rm g}^{-1}\right]_{_{\rm CO.LOW}}\,\rho^2\,a^3,
\end{eqnarray}
where we have assumed that the fractional abundance of $^{12}$C$^{16}$O relative to H$_2$, by number, is $3\times 10^{-5}$.

If the density is high, the level populations are thermalised by frequent collisions, and photon emissions account for only a small fraction of de-excitations. If the lines are also optically thick, then the spectrum can be approximated by a blackbody filtered through the discrete lines where the gas has opacity, and therefore emissivity. The net frequency width of a blackbody spectrum is $\Delta\nu_{_{\rm BB}}\!\sim\! f_{_{\rm CONT}}k_{_{\rm B}}T/h$, with $f_{_{\rm CONT}}\!\sim\! 5$. The net frequency width of the discrete rotational lines from a single linear molecule is
\begin{eqnarray}\nonumber
\Delta\nu_{_{\rm MOL}}&\simeq&f_{_{\rm LINE}}\,\sum\limits_{J=1}^{J=J_{_{\rm MAX}}}\left\{\frac{(E_{_J}-E_{_{J-1}})}{h}\right\}\,\frac{\Delta v}{c}\\\nonumber
&\simeq&f_{_{\rm LINE}}\,\int\limits_{J=0}^{J=J_{_{\rm MAX}}}\frac{J\hbar^2}{hI}\,dJ\,\frac{\Delta v}{c}\\
&\simeq&\frac{f_{_{\rm LINE}}k_{_{\rm B}}T}{h}\,\frac{\Delta v}{c}\,.
\end{eqnarray}
Here, $\Delta v$ is the velocity dispersion in the gas (which is presumed to determine the individual line-widths, i.e. we are not in the realm of natural or pressure broadening), and converting the summation to an integral is appropriate as long as $J_{_{\rm MAX}}\!\gg\!1$. $\;f_{_{\rm LINE}}$ is a measure of (a) the extent to which high-energy levels are excited, and (b) the extent to which the equivalent widths of emission lines exceed $\Delta v$. For $^{12}$C$^{16}$O we adopt $f_{_{\rm LINE}}\!\sim\!20$, to match the results of Goldsmith \& Langer (1978). It follows that the net cooling rate due to $^{12}$C$^{16}$O, in the high-density, optically thick regime, is
\begin{eqnarray}\nonumber
\Lambda_{_{\rm CO.HIGH}}\!\!&\!\!\sim\!\!&\!\!\frac{4\sigma_{_{\rm SB}}T^4}{L}\frac{\Delta\nu_{_{\rm MOL}}}{\Delta\nu_{_{\rm BB}}}\\\nonumber
\!\!&\!\!\sim\!\!&\!\!\frac{4\sigma_{_{\rm SB}}T^4}{c}\frac{f_{_{\rm LINE}}}{f_{_{\rm CONT}}}\frac{\Delta v}{L}\\\label{EQN:DENSE;THICK1}
\!\!&\!\!\sim\!\!&\!\!\left[1.7\times 10^{-44}\,{\rm cm}^{-9}\,{\rm g}\,{\rm s}^6\right]_{_{\rm CO.HIGH}}\,a^8\,\frac{\Delta v}{L}\,,
\end{eqnarray}
Here $L$ is the linear size of the emitting region, and the combination $\Delta v/L$ is equivalent to $|dv/dR|$ in the analysis of \citet{GoldLang1978}. As pointed out by \citet{GoldKwan1974}, expressions for the cooling rate in the dense, optically thick limit depend only weakly on the properties of the molecule concerned, principally via the level spacing, provided that the species exists in the gas phase, and has rotational levels that can be excited collisionally at the densities and temperatures involved.

We can obtain an approximate fit to the net cooling due to $^{12}$C$^{16}$O by combining Eqs. (\ref{EQN:RAREFIED;THIN}) and (\ref{EQN:DENSE;THICK1}) thus:
\begin{eqnarray}\label{EQN:NET1}
\Lambda_{_{\rm CO.TOT}}&\simeq&\left\{\Lambda_{_{\rm CO.LOW}}^{-1}+\Lambda_{_{\rm CO.HIGH}}^{-1}\right\}^{-1}.
\end{eqnarray}
This expression somewhat overestimates the cooling rate in the intermediate regime where the level populations are starting to thermalise and the transitions are starting to become optically thick. However, this overestimate does not affect our conclusion that $^{12}$C$^{16}$O is a very effective post-shock coolant (Section \ref{SEC:SHOCK_CO}), and it strengthens our conclusion that $^{12}$C$^{16}$O is not an important coolant in collapsing prestellar cores (Section \ref{SEC:CORE_CO}).

\section{Dust cooling}\label{APP:DUST}

Cooling by dust involves two stages. In the first stage the thermal energy of the gas is transferred to the dust. In the second stage the dust radiates the energy away into space. Although the energy transferred from the gas to the dust is very important for the thermal balance of the gas, it is normally trivial for the thermal balance of the dust, under the circumstances with which we are concerned here (see below). Thus the rate of gas cooling due to dust, per unit volume, is controlled by the rate at which thermal energy is transferred from the gas to the dust (subscript {\tiny G2D}),
\begin{eqnarray}\nonumber
\Lambda_{_{\rm G2D}}&\sim&n_{_{\rm TOT}}\,\bar{v}(T)\,n_{_{\rm DUST}}\,\sigma_{_{\rm DUST}}\,\frac{3k_{_{\rm B}}(T-T_{_{\rm DUST}})}{2}\\\nonumber
&&\hspace{4.0cm}\times\,\alpha_{_{\rm DUST}}(T,T_{_{\rm DUST}})\\\nonumber
&\sim&3\left(\frac{2}{\pi}\right)^{1/2}\,\kappa\subD\;\Delta{\rm ln}[T]_{_{\rm G2D}}\,\rho^2\,a^3\\\label{EQN:G2D1}
&\sim&\left[1.2\times 10^2\,{\rm cm}^2\,{\rm g}^{-1}\right]_{_{\rm G2D}}\,\rho^2\,a^3\,.
\end{eqnarray}
On the first line of Eq. (\ref{EQN:G2D1}), $n_{_{\rm TOT}}$ is the total number-density of gas particles, $\bar{v}(T)\!\simeq\!(8k_{_{\rm B}}T/\pi\bar{m})^{1/2}$ is their arithmetic mean speed, $n_{_{\rm DUST}}$ is the number-density of dust particles, and $\sigma_{_{\rm DUST}}$ is the geometric cross-section of a dust grain. $\;3k_{_{\rm B}}(T-T_{_{\rm DUST}})/2$ is the mean thermal energy that is lost from the gas if a gas particle becomes adsorbed on the grain surface, comes into thermal equilibrium with the grain, and is subsequently thermally evaporated from the grain surface with a speed corresponding to the grain temperature, $T_{_{\rm DUST}}$. $\;\alpha_{_{\rm DUST}}(T,T_{_{\rm DUST}})$ is a factor to correct for the fact that the energy loss will be reduced if the gas particle does not reach complete thermal equilibrium with the grain (for example, it may simply bounce off the grain) and for other factors (the gas particle may be released from the grain surface by a cosmic ray or by a thermal spike due to stochastic heating of a small grain by an individual UV photon, the effective cross-section of the grain may be altered if the grain and/or the particle are charged, and so on). In the temperature regime with which we are concerned it is normally presumed that $\alpha_{_{\rm DUST}}(T,T_{_{\rm DUST}})\sim 1$ \citep[e.g.][]{HollMcKe1979}, and we have done so too. 

To obtain the second line of Eq. (\ref{EQN:G2D1}) we have substituted $\,n_{_{\rm DUST}}\sigma_{_{\rm DUST}}\!=\!\rho\kappa\subD$ (see Eq. \ref{EQN:kappaD1}) and $\;\Delta{\rm ln}[T]_{_{\rm G2D}}\!\equiv\!(T-T_{_{\rm DUST}})/T$. We then stipulate that thermal coupling requires $\,\Delta{\rm ln}[T]_{_{\rm G2D}}\la 0.2$, i.e. $\,T_{_{\rm DUST}}\la T\la 1.2\,T_{_{\rm DUST}}$; thus, for example, if $\,T_{_{\rm DUST}}=8\,{\rm K}$, the gas is only deemed to be coupled thermally to the dust if $\Lambda_{_{\rm G2D}}$ can deliver $\,8\,{\rm K}<T\la 10\,{\rm K}$. 

To obtain the third line of Eq. (\ref{EQN:G2D1}) we have therefore substituted $\,\Delta{\rm ln}[T]_{_{\rm G2D}}= 0.2$. If in reality $\;\Delta{\rm ln}[T]_{_{\rm G2D}}\!>\!0.2$, we are underestimating $\Lambda_{_{\rm G2D}}$, which should make our conclusions stronger -- except that, if this is the case, the thermal balance of the gas is probably not significantly affected by coupling to the dust anyway. Conversely, if in reality $\;\Delta{\rm ln}[T]_{_{\rm G2D}}\!<\!0.2$, we are overestimating $\Lambda_{_{\rm G2D}}$, but because of the $\rho^2$ term, this will only matter in situations where the smallness of $\;\Delta{\rm ln}[T]_{_{\rm G2D}}$ reflects how strongly the gas is thermally coupled to the dust; under this circumstance, $\;\Delta{\rm ln}[T]_{_{\rm G2D}}$ adjusts so that $\Lambda_{_{\rm G2D}}$ balances the net heating from other sources (normally mainly compression).

Once the thermal energy of the gas has been transferred to the dust, it must be radiated away. However, under the circumstances with which we are concerned, this occurs very quickly. In effect the dust acts like a thermodynamic heat reservoir for the gas. This appears paradoxical, since the dust constitutes only a small fraction of the mass ($Z_{_{\rm DUST}}\!\sim\!0.01$), and therefore its thermal capacity is much less than that of the gas. However, the thermal balance of the dust is dominated by the absorption and emission of continuum radiation, and the resulting energy turnover, i.e. the rate of absorption and emission of radiation by a dust grain, is so high that the thermal energy transferred to the grain from the gas is a small perturbation and has almost no influence on the temperature of the dust. To see this, the rate per unit volume at which the dust radiates (subscript {\tiny D2R}) is
\begin{eqnarray}\nonumber
\Lambda_{_{\rm D2R}}&\sim&n_{_{\rm DUST}}\,4\sigma_{_{\rm DUST}}\,\sigma_{_{\rm SB}}T_{_{\rm DUST}}^4\;{\bar Q}_{_{\rm PLANCK}}(T_{_{\rm DUST}})\\\nonumber
&\sim&\frac{8\,\pi^5\,{\bar m}^6\,\kappa\subO\,{\bar Q}\subO}{15\,c^2\,h^3\,(k_{_{\rm B}}{\rm K})^2}\,\left(\!\frac{T_{_{\rm DUST}}}{T}\!\right)^6\,\rho\,a^{12}\\\label{EQN:DUST2RAD}
&\sim&\left[7\times 10^{-53}\,{\rm cm}^{-10}\,{\rm s}^9\right]_{_{\rm D2R}}\left(\!\frac{T_{_{\rm DUST}}}{T}\!\right)^6\,\rho\,a^{12}\,.
\end{eqnarray}
Here, $4\sigma_{_{\rm DUST}}\!=\!4\pi r_{_{\rm DUST}}^2$ is the surface area of a grain, $\sigma_{_{\rm SB}}T_{_{\rm DUST}}^4$ is the blackbody flux from a surface at temperature $T_{_{\rm DUST}}$, $\bar{Q}_{_{\rm PLANCK}}(T_{_{\rm DUST}})$ is the Planck-mean emission efficiency of the dust at this temperature (see Eq. \ref{EQN:QPlanck}), and we are assuming that the dust grains are  in thermal equilibrium (which is a good assumption for the grains inside a prestellar core, since there are very few short-wavelength photons around that could heat transiently any small dust grains still present). With $\Delta{\rm ln}[T]_{_{\rm G2D}}\!<\!0.2$ we have $0.26\la \left(T_{_{\rm DUST}}/T\right)^6<1$. Thus at the temperatures and densities with which we are concerned here (see Section \ref{SEC:BASIC}), the fractional contribution to dust heating from the gas is 
\begin{eqnarray}\nonumber
\frac{\Lambda_{_{\rm G2D}}}{\Lambda_{_{\rm D2R}}}\!\!&\!\!\la\!\!&\!\!\frac{0.20\,[{\rm Eq\,E1}]_{_{\rm G2D}}}{0.26\,[{\rm Eq\,E2}]_{_{\rm D2R}}}\;\rho\,a^{-9}\\\label{EQN:G2D2D2R}
&\la&6.5\times 10^{-6}\,\left(\frac{n\subH}{100\,{\rm cm}^{-3}}\right)\,\left(\frac{T}{10\,{\rm K}}\right)^{\!-9/2}.
\end{eqnarray}
We conclude that, {\it in the density and temperature regime with which we are concerned}, the gas makes a very small contribution to the heating of the dust. However, as the subsequent condensation of a prestellar core proceeds and the density and column-density rise, this contribution increases, more or less monotonically, and eventually it becomes dominant.

\section{Relative timescales}\label{APP:TIMESCALES}

We obtain the ratio of the CO cooling timescale, $3\rho a^2/2\Lambda_{_{\rm CO.TOT}}$, to the freefall timescale, by substituting from Eqns. (\ref{EQN:CO_TOT1}), (\ref{EQN:CO_LOW1}), (\ref{EQN:CO_HIGH3}) and (\ref{EQN:FREEFALL}), 
\begin{eqnarray}\nonumber
\frac{t_{_{\rm CO\,COOLING}}}{t_{_{\rm FREEFALL}}}&\sim&0.004\,\left(\frac{n_{_{{\rm H}_2}}}{10^5\,{\rm cm}^{-3}}\right)^{-1/2}\,\left(\frac{T}{10\,{\rm K}}\right)^{-1/2}\\
&&\hspace{0.4cm}+\;4.3\,\left(\frac{n_{_{{\rm H}_2}}}{10^5\,{\rm cm}^{-3}}\right)\,\left(\frac{T}{10\,{\rm K}}\right)^{-3};
\end{eqnarray}
the first term on the left represents the low-density, optically thin regime, and the second term represents the high-density optically thick regime.

The timescale for freeze-out of CO is given by
\begin{eqnarray}\nonumber
t_{_{\rm CO\,FREEZEOUT}}&=&\frac{1}{\bar{v}_{_{\rm CO}}\,n\subD\,\sigma\subD}\\
&=&\frac{1}{(8k_{_{\rm B}}T/\pi m_{_{\rm CO}})^{1/2}\,\rho\,\kappa\subD}\,,
\end{eqnarray}
where $\bar{v}_{_{\rm CO}}=(8k_{_{\rm B}}T/\pi m_{_{\rm CO}})^{1/2}$ is the arithmetic mean speed of a CO molecule. Combining this with Eqn. (\ref{EQN:FREEFALL}), we obtain
\begin{eqnarray}
\frac{t_{_{\rm CO\,FREEZEOUT}}}{t_{_{\rm FREEFALL}}}\!&\!\sim\!&\!0.34\,\left(\!\frac{n_{_{{\rm H}_2}}}{10^5\,{\rm cm}^{-3}}\!\right)^{\!-1/2}\,\left(\!\frac{T}{10\,{\rm K}}\!\right)^{\!-1/2}\!.
\end{eqnarray}

Finally, we obtain the ratio of the dust cooling timescale, $3\rho a^2/2\Lambda_{_{\rm G2D}}$, to the freefall timescale, by substituting from Eqns. (\ref{EQN:G2D1}) and (\ref{EQN:FREEFALL}),
\begin{eqnarray}
\frac{t_{_{\rm DUST\,COOLING}}}{t_{_{\rm FREEFALL}}}\!&\!\sim\!&\!0.43\,\left(\!\frac{n_{_{{\rm H}_2}}}{10^5\,{\rm cm}^{-3}}\!\right)^{\!-1/2}\,\left(\!\frac{T}{10\,{\rm K}}\!\right)^{\!-1/2}\!.
\end{eqnarray}

The logarithms of these ratios are plotted against $\log_{_{10}}(n_{_{{\rm H}_2}}/{\rm cm}^{-3})$ on Fig. \ref{FIG:Cooling}, as a thick dashed line ($t_{_{\rm CO\,COOLING}}/t_{_{\rm FREEFALL}}$), a thick dotted line ($t_{_{\rm CO\,FREEZEOUT}}/t_{_{\rm FREEFALL}}$) and a thick full line ($t_{_{\rm DUST\,COOLING}}/t_{_{\rm FREEFALL}}$). A thin dotted horizontal line at $\log_{_{10}}(t/t_{_{\rm FREEFALL}})\!=\!0$ is included for reference. CO cooling and dust cooling can only support collapse and fragmentation where the corresponding lines ({\sc co cooling} and {\sc dust cooling}) are below this horizontal line; conversely CO freezes out onto the dust where the {\sc co freezeout} line is below the horizontal line, and once frozen out it will not contribute to cooling, except in as much as it makes the dust grains a bit larger.

\bibliographystyle{mn2e}
\bibliography{antsrefs}

\label{lastpage}
\end{document}